# Modelling particle collisions in moderately dense curtain impacted by an incident shock wave

Pikai Zhang[1,2], Huangwei Zhang[1,2 1], Yun Feng Zhang[1,2], Shangpeng Li[1], and Qingyang Meng[2]

[1] Department of Mechanical Engineering, National University of Singapore, 9 Engineering Drive 1, Singapore 117576, Republic of Singapore
[2] National University of Singapore (Chongqing) Research Institute, Liangjiang New Area, Chongqing 401123, People's Republic of China

**ABSTRACT**

The interactions between an incident shock and moderately dense particle curtain are simulated with the Eulerian-Lagrangian method. A customized solver based on OpenFOAM is extended with an improved drag model and collision model, and then validated against two benchmark experiments. In this work, parametric studies are performed considering different particle sizes, volume fractions, and curtain thicknesses. It is found that smaller particle size and larger volume fractions lead to stronger reflected shock and weaker transmitted shock. Different expansion stages of the curtain fronts are also studied in detail. Attention is paid to the particle collision effects on the curtain evolution behaviours. According to our results, for the mono-dispersed particle curtain, the collision effects on curtain front behaviors are small, even when the initial particle volume fraction is as high as 20%. This is due to the positive velocity gradient across the curtain after the shock wave passage, leading to faster motion of downstream particles than the upstream ones and hence no collision occurs. For the bi-dispersed particle curtain, the collision effects become important in the mixing region of different-size particles. Collisions decelerate small particles while accelerate large ones and cause velocity scattering. Moreover, increasing the bi-dispersed curtain thickness leads to multiple collision force peaks due to the local particle accumulations, which is the result of the delayed separation of different particle groups. Our results indicate that the collision model may be unnecessary to predict curtain fronts in mono-dispersed particles, but in bi-dispersed particles, the collision effects are important and therefore must be modelled.

[1] Corresponding author. Email: Huangwei.zhang@nus.edu.sg.

**Nomenclature**

| | |
|---|---|
| $A_d$ | Surface area of one particle [$m^2$] |
| $\mathbf{a}_\mathrm{L}$ | Curtain expansion acceleration [$m/s^2$] |
| $A_s$ | Sutherland coefficient |
| $c$ | Sound speed [m/s] |
| $c_{p,g}$ | Gas heat capacity at constant pressure [J/kg/K] |
| $c_{p,d}$ | Particle heat capacity at constant pressure [J/kg/K] |
| $C_1$ | Drag correction function |
| $C_D$ | Standard drag coefficient |
| $C_{D,correction}$ | Correction drag coefficient |
| $\mathbf{D}$ | Deformation gradient tensor |
| $D_d$ | Particle diameter [m] |
| $E$ | Total non-chemical energy [J/kg] |
| $e$ | Specific sensible internal energy [J/kg] |
| $\mathbf{e}_\mathrm{k}$ | Unit vector of $k$-direction |
| $\mathbf{F}_\mathrm{pg}$ | Pressure gradient force [N] |
| $\mathbf{F}_\mathrm{pp}$ | Collision force [N] |
| $\mathbf{F}_\mathrm{qs}$ | Quasi-steady force [N] |
| $\mathbf{F}_\mathrm{surf}$ | Fluid dynamic force [N] |
| $h_c$ | Convective heat transfer coefficient [$W/m^2/K$] |
| $\mathbf{I}$ | Unit tensor |
| $k_c$ | Thermal conductivity coefficient [W/m/K] |
| $L$ | Particle curtain thickness [m] |
| $m$ | Particle number density [1/m] |
| $m_d$ | Mass of a single particle [kg] |
| $Ma_d$ | Particle Mach number |
| $N_i$ | Particle number in one cell |
| $Nu$ | Nusselt number |
| $p$ | Pressure [Pa] |
| $P_s$ | Pressure constant [Pa] |
| $Pr$ | Prandtl number |
| $\mathbf{q}$ | Diffusive heat flux [$W/m^2$] |
| $\dot{Q}_c$ | Convective heat transfer rate [J/s] |
| $R$ | Specific gas constant [J/kg/K] |
| $Re_d$ | Particle Reynolds number |
| $S_{energy}$ | Energy source term [$J/m^3/s$] |
| $\mathbf{S}_\mathrm{mom}$ | Momentum source term [$N/m^3$] |
| $t$ | Time [s] |
| $T_c$ | Gas temperature [K] |
| $T_d$ | Particle temperature [K] |
| $T_s$ | Sutherland temperature [K] |
| $\mathbf{u}_\mathrm{c}$ | Gas velocity [m/s] |
| $\mathbf{u}_\mathrm{d}$ | Particle velocity [m/s] |
| $\bar{\mathbf{u}}_\mathrm{d}$ | Particle intermediate velocity [m/s] |
| $\tilde{\mathbf{u}}_\mathrm{d}$ | Mass-averaged particle velocity [m/s] |
| $\mathbf{U}_\mathrm{L}$ | Curtain expansion velocity [m/s] |
| $\mathbf{u}_{\mathrm{p}\tau}$ | Collision correction velocity [m/s] |
| $\mathbf{U}_\mathrm{s}$ | Shock wave velocity [m/s] |
| $\mathbf{U}_{\Delta\mathrm{p}}$ | Pressure-based velocity [m/s] |
| $V_c$ | CFD cell volume [$m^3$] |
| $V_d$ | Particle volume [$m^3$] |
| $x_d$ | Location of individual particle [m] |
| $Z$ | Acoustic impedance [$kg/m^2/s$] |

***Greek letters***

| | |
|---|---|
| $\alpha$ | Volume fraction [%] |
| $\mu$ | Dynamic viscosity [kg/m/s] |
| $\rho$ | Density [kg/m$^3$] |
| $\tau$ | Solid stress [Pa], response time [s] |
| $\boldsymbol{\tau}$ | Viscous stress tensor [Pa] |
| $\gamma$ | Heat capacity ratio, restitution coefficient |
| $\Delta$ | Mesh size [m] |

***Superscripts***

| | |
|---|---|
| T | Transpose |
| $\beta$ | Constant value |
| 0 | Initial value |
| n | Time step |
| * | Dimensionless quantity |

***Acronym***

| | |
|---|---|
| CFD | Computation fluid dynamics |
| CMF | Compressible multiphase flows |
| CAR | Constant acceleration regime |
| CTR | Constant thickness regime |
| CVR | Constant velocity regime |
| DCF | Downstream curtain front |
| E-E | Eulerian-Eulerian |
| E-L | Eulerian-Lagrangian |
| KNP | Kurganov, Noelle and Petrova |
| MP-PIC | Multiphase Particle-In-Cell |
| ODE | Ordinary differential equations |
| UCF | Upstream curtain front |

***Subscripts***

| | |
|---|---|
| $c$ | Continuous phase, convective |
| $curtain$ | Particle curtain |
| $coll$ | Collision |
| $d$ | Dispersed phase, drag |
| $D, std$ | Standard drag coefficient |
| $D, sub$ | Subcritical drag coefficient |
| $D, sup$ | Supercritical drag coefficient |
| $e$ | Equilibrium |
| $energy$ | Energy |
| mom | Momentum |
| $p$ | Particle |
| $pl$ | Packing volume limit |
| pg | Pressure gradient |
| pp | Particle-particle collision |
| pτ | Collision correction |
| qs | Quasi-steady |
| surf | Surface |
| $V$ | Momentum response |
| 0 | Initial state |
| 1, 2 | Medium species |

## 1. INTRODUCTION

Compressible Multiphase Flows (CMF), containing shock and particles/droplets, exist in a variety of engineering systems, e.g., cold spray [1], supersonic combustion [2] and high energy explosions [3]. Most CFM systems are usually characterized by strongly varying dispersed phase volume fraction, ranging from dense to dilute loading. Two-way coupling is sufficient for dilute flows, but four-way coupling, that is, including particle-particle interactions, must be considered when the CMF is dense. Figure 1 summarizes previous experimental and numerical studies on shock-particle curtain interactions. It is shown that the considered particle diameters are mainly 50 – 500 μm, while the particle volume fraction is basically distributed between 4% – 40%, which falls into the region of collision- and contact-dominated regimes, suggesting that the particle-particle interactions may play a significant role.

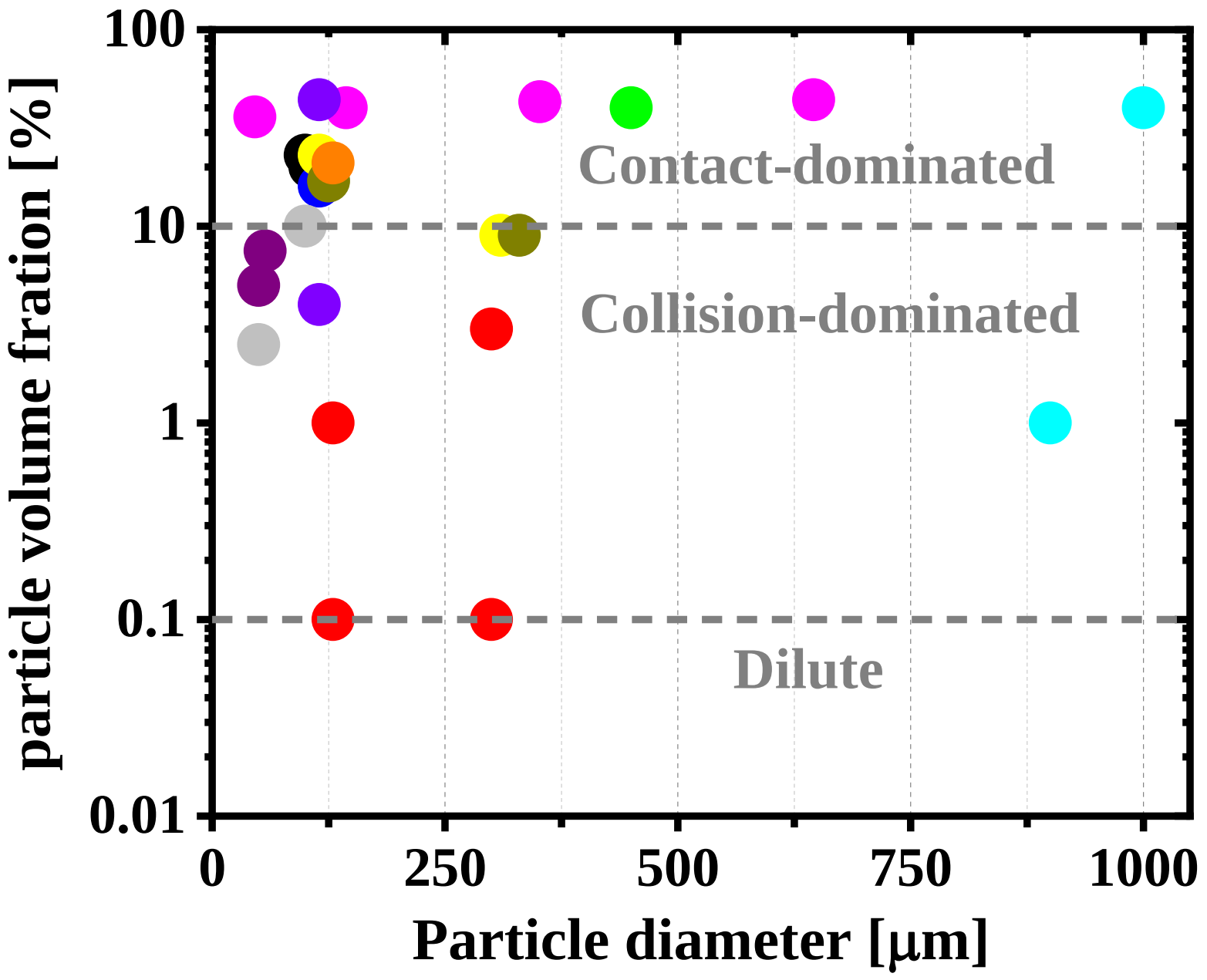

Fig. 1 Summary of particle curtain properties in different compressible gas-particle flows. ●: Wagner et al. [4, 5], ●: Boiko et al. [6], ●: Sugiyama et al. [7], ●: Ling et al. [8]; ●: Park et al. [9], ●: Theofanous [10, 11], ●: Lv et al. [12], ●: DeMauro et al. [13, 14], ●: Daniel et al. [15], ●: Mehta et al. [16], ●: Osnes et al. [17], ●: Shallcross et al. [18]. Horizontal lines: the boundaries between dilute, collision-dominated, and contact-dominated flows [19].

For the gas phase, when an incident shock wave impacts a dense particle curtain, a reflected shock forms and travels back upstream [4]. The strength of the reflected shock is positively correlated with initial particle volume fraction [17] and curtain thickness [12], whilst negatively correlated with particle size [20] and particle density [15]. When the post-incident-shock flow is supersonic, Boiko et al. [6] observe a bow shock in front of the particle cloud with the initial volume fraction being inside the collision-dominated regime as shown in Fig. *1*.

For the particulate phase, moderately dense particle concentration typically results in different curtain morphological evolutions from those of dilute curtains. Wagner et al. [4], for instance, reveal the behaviours of upstream and downstream fronts of the curtain. Tens of microseconds after the incident shock impingement, the downstream particles start to move, and after a period the upstream ones move at a lower velocity, resulting in curtain expansion. Ling et al. [8] further explain that the curtain expansion is mainly caused by the positive gas velocity gradient and negative pressure gradient inside the curtain. For the flows in the contact-dominated regime (see Fig. *1*), Theofanous et al. [10] and DeMauro et al. [14] use scaling laws to describe the three stages of curtain expansion: constant-thickness regime (CTR), constant-acceleration regime (CAR) and constant-velocity regime (CVR). Furthermore, Wagner et al. [5] measure the time evolutions of the volume fraction profile as the curtain expands and find that the downstream side is featured by higher volume fractions. Nevertheless, Theofanous et al. [11] discover that such high volume fraction would move to the upstream side when the post-incident-shock flow increases to supersonic condition.

When the foregoing particle behaviours are analysed, limited studies have been conducted to understand the force of the particle motion. Sugiyama et al. [7] numerically studied the drag force and pressure gradient force of particles in the contact-dominated regimes (see Fig. *1*). They find the drag force remains almost constant, while the pressure gradient force becomes small as the particle curtain expands. Moreover, Mehta et al. [16] calculate the drag force histories of each particle, and find the multiple-particle system alters the drag correlation of each particle, which means that corrections are needed to the single-particle drag model.

Particle-particle interactions have long been considered to be important in the dense particle regime. Although many previous simulations of dense flows have incorporated the collision models [7, 8, 21, 22], their effects have not been studied in detail. In this work, the particle collisions within a moderately dense curtain will be studied based on a Eulerian-Lagrangian approach with four-way coupling, which considers both fluid-particle interactions and particle-particle collisions. Specifically, we aim to: (1) extend the existing compressible two-phase flow solver *RYrhoCentralFoam* [23, 24] to model moderately dense flows; (2) reveal the roles of inter-particle collisions in both mono- and bi-dispersed curtains; (3) analyze the fluid dynamic force and collision force resulting in various curtain evolution behaviors. In the rest of the manuscript, the governing equations, numerical implementations, and validations are detailed in sections 2-4, respectively. The physical model of shock-curtain interactions is given in section 5. The results and discussion are presented in section 6, followed by the conclusions in section 7.

## 2. GOVERNING EQUATIONS

To model dense gas-particle two-phase flows, two approaches have been developed. In the Eulerian-Eulerian (E-E) method (macroscopic descriptions), Baer and Nunziato include the compressibility of gas and particle phases and the compaction behaviors of the dense mixture [25]. However, a non-disturbing condition imposes numerical challenges in solving the coupled system [26]. This is related to a new nozzling term, accounting for the gas acceleration due to the flow area restriction by particles. For this reason, Houim and Oran recently design the algorithms to accurately discretize the coupled gas and granular Euler equations in the E-E method [27].

The Eulerian-Lagrangian (E-L) method (microscopic description) can capture the behaviours of individual particles, and hence provide particle-scale information (e.g., force and trajectory). The computational cost of the E-L method dramatically increases when large number of particles and their interactions are considered [28]. In this situation, the *computational parcel* concept [29] can be introduced to reduce the overhead, which groups the particles with identical size, velocity, and position.

Panchal and Menon [30] recently propose a hybrid E-E and E-L method for two-phase flows with dilute to dense particle loading. A transition criterion and algorithm for the conversion between the Lagrangian and Eulerian descriptions for the dispersed phase are developed. However, their work is limited to mono-dispersed particles. Moreover, Tian et al. [31] extend the E-L method with coarse-grained DEM approach to model compressible dense and even granular flows. However, interphase heat transfer is absent in their framework and therefore the temperature of gas and particle phases may not be predicted accurately.

Based on OpenFOAM, we develop a comprehensive E-L framework for simulating two-phase compressible reacting flows [23] and have employed it for various two-phase detonation or supersonic combustion problems [2, 32-34]. Nonetheless, it is designed only for dilute two-phase flows with particle volume fraction $\alpha_d$ < 0.1%, and hence the volume fraction effects on the evolutions of gas phase quantities are neglected. In this work, we will extend our current E-L framework for simulating moderately dense two-phase flows ($\alpha_d$ up to 10-20% [19]), and the formulations are presented below.

### 2.1 Gas phase

The volume fractions of the particle ($\alpha_d$) and gas ($\alpha_c$) phases in a CFD cell are

$$\alpha_d \equiv \frac{\sum_{N_i} V_d}{V_c} \quad \text{and} \quad \alpha_c = 1 - \alpha_d. \tag{1}$$

Here $N_i$ is the particle number in the cell, $V_d$ and $V_c$ are the volume of the particle and cell, respectively. The particle volume fraction effects should be considered for dense two-phase flows, and the governing equations of mass, momentum, and energy of the gas phase read [19, 35]

$$\frac{\partial(\alpha_c \rho_c)}{\partial t} + \nabla \cdot (\alpha_c \rho_c \mathbf{u}_\mathrm{c}) = 0, \tag{2}$$

$$\frac{\partial(\alpha_c \rho_c \mathbf{u}_\mathrm{c})}{\partial t} + \nabla \cdot (\alpha_c \rho_c \mathbf{u}_\mathrm{c} \mathbf{u}_\mathrm{c}) = -\nabla p + \nabla \cdot \boldsymbol{\tau} + \mathbf{S}_\mathrm{mom}, \tag{3}$$

$$\begin{aligned} \frac{\partial(\alpha_c \rho_c E)}{\partial t} + \nabla \cdot [(\alpha_c \rho_c E + \alpha_c p)\mathbf{u}_\mathrm{c}] = -\nabla \cdot (\alpha_d p \tilde{\mathbf{u}}_\mathrm{d}) + \nabla \cdot (\alpha_c \boldsymbol{\tau} \cdot \mathbf{u}_\mathrm{c}) + \\ \nabla \cdot (\alpha_d \boldsymbol{\tau} \cdot \tilde{\mathbf{u}}_\mathrm{d}) - \nabla \cdot (\alpha_c \mathbf{q}) + S_{energy}. \end{aligned} \tag{4}$$

$t$ is time, $\rho_c$ the gas density, and $\mathbf{u}_\mathrm{c}$ the gas velocity vector. The pressure $p$ is calculated from the ideal gas equation of state

$$p = \rho_c R T_c, \tag{5}$$

where $R$ is the specific gas constant and $T_c$ is the gas temperature. The total energy $E$ is

$$E \equiv e + \frac{\mathbf{u}_\mathrm{c}^2}{2}, \tag{6}$$

where $e$ is the specific internal energy.

In Eq. (4), $\tilde{\mathbf{u}}_\mathrm{d}$ is the mass-averaged particle velocity in individual CFD cells, i.e.,

$$\tilde{\mathbf{u}}_\mathrm{d} = \frac{\sum_{N_i}(\rho_d V_d \mathbf{u}_\mathrm{d})}{\sum_{N_i}(\rho_d V_d)}, \tag{7}$$

in which $\rho_d$ and $\mathbf{u}_\mathrm{d}$ are the particle material density and velocity, respectively. The viscous stress tensor $\boldsymbol{\tau}$ in the momentum equation, i.e., Eq. (3), is modelled as

$$\boldsymbol{\tau} = -2\mu_c \mathrm{dev}(\mathbf{D});\ \mathrm{dev}(\mathbf{D}) = \mathbf{D} - \frac{\mathrm{tr}(\mathbf{D})\mathbf{I}}{3};\ \mathbf{D} = \frac{[\nabla\mathbf{u}_\mathrm{c} + (\nabla\mathbf{u}_\mathrm{c})^\mathrm{T}]}{2}, \tag{8}$$

where $\mu_c$ is the dynamic viscosity estimated with the Sutherland law [36], i.e., $\mu_c = A_s\sqrt{T}/(1 + T_s/T)$. Here $A_s$ is the Sutherland coefficient and $T_s$ is the Sutherland temperature. **I** is the unit tensor, **D** the deformation gradient tensor, and $\mathrm{dev}(\mathbf{D})$ its deviatoric component.

In Eq. (4), $\mathbf{q}$ is the diffusive heat flux, calculated by the Fourier's law

$$\mathbf{q} = -k_c \nabla T_c, \tag{9}$$

where $k_c$ is the thermal conductivity of the gas phase.

The order of magnitude of two particle-related viscous work terms can be compared as

$$\frac{\nabla\cdot(\alpha_d \boldsymbol{\tau}\cdot\tilde{\mathbf{u}}_\mathrm{d})}{\nabla\cdot(\alpha_d p \tilde{\mathbf{u}}_\mathrm{d})} \sim \frac{\mu_c}{\rho_c \mathbf{U} L} = \frac{1}{Re_d}, \tag{10}$$

in which **U** and $L$ are the characteristic velocity and length scales and estimated with $\mathbf{u}_\mathrm{c} - \mathbf{u}_\mathrm{d}$ and $D_d$, respectively. $D_d$ is the particle diameter. The particle Reynolds number is $Re_d \equiv \rho_c D_d |\mathbf{u}_\mathrm{c} - \mathbf{u}_\mathrm{d}|/\mu_c$. Since $Re_d \gg 1$ is generally valid for compressible flows, the work by the particle viscous force, i.e., $\nabla \cdot (\alpha_d \boldsymbol{\tau} \cdot \tilde{\mathbf{u}}_\mathrm{d})$, is neglected in this study.

The source terms in Eqs. (3) and (4), $\mathbf{S}_\mathrm{mom}$ and $S_{energy}$, account for the momentum and energy exchange between two phases

$$\mathbf{S}_\mathrm{mom} = -\frac{1}{V_c}\sum_{N_i}\mathbf{F}_\mathrm{surf}, \tag{11}$$

$$S_{energy} = -\frac{1}{V_c}\sum_{N_i}(\mathbf{F}_\mathrm{surf}\cdot\mathbf{u}_\mathrm{d} + \dot{Q}_c). \tag{12}$$

$\mathbf{F}_{\text{surf}}$ and $\dot{Q}_c$ are the fluid dynamic force and convective heat transfer, respectively. Their expressions will be given in the next section.

## 2.2 Particle phase

The Lagrangian method is adopted to track the individual particles in the dispersed phase. Point-force approximation is used, which means that the particles are not resolved and are represented by points with mass. The particles are assumed to be spherical. Meanwhile, the particle Biot number is generally small and therefore the particle temperature is assumed to be uniform. The evolutions of particle momentum and energy are governed by

$$m_d \frac{\mathrm{d}\mathbf{u}_\mathrm{d}}{\mathrm{d}t} = \mathbf{F}_{\text{surf}} + \mathbf{F}_{\text{pp}}, \tag{13}$$

$$m_d c_{p,d} \frac{\mathrm{d}T_d}{\mathrm{d}t} = \dot{Q}_c. \tag{14}$$

Here $m_d = \pi \rho_d D_d^3/6$ is the particle mass, $T_d$ the particle temperature, and $c_{p,d}$ the particle heat capacity. The convective heat transfer rate $\dot{Q}_c$ in Eq. (14) is

$$\dot{Q}_c = h_c A_d (T_c - T_d), \tag{15}$$

where $A_d = \pi D_d^2$ is the surface area of a particle. $h_c$ is the convective heat transfer coefficient, computed with the Ranz and Marshall correlation [37]

$$Nu = \frac{h_c D_d}{k_c} = 2.0 + 0.6 Re_d^{1/2} Pr^{1/3}, \tag{16}$$

where $Pr = \mu_c c_{p,g}/k_c$ is the gas Prandtl number, and $c_{p,g}$ is the heat capacity of gas phase at constant pressure.

The fluid dynamic force, $\mathbf{F}_{\text{surf}}$, includes different components

$$\mathbf{F}_{\text{surf}} = \mathbf{F}_{\text{qs}} + \mathbf{F}_{\text{pg}}, \tag{17}$$

in which $\mathbf{F}_{\text{qs}}$ is the quasi-steady force. $\mathbf{F}_{\text{pg}}$ is the pressure gradient force, calculated with $\mathbf{F}_{\text{pg}} = -V_d \nabla p$. Other forces (e.g., added-mass force and Basset force) are safely ignored considering their secondary significance [22]. In Eq. (13), $\mathbf{F}_{\text{pp}}$ is the collision force and will be introduced in section 2.3.

In compressible dense two-phase flows, the effects of particle Reynolds number, volume fraction, and Mach number must be considered when the particle drag force $\mathbf{F}_{\rm qs}$ is modelled [19, 38]. Gidaspow [39] proposes a drag model, i.e.,

$$\mathbf{F}_{\rm qs} = \begin{cases} \frac{m_d \mu_c (150\alpha_d + 1.75\alpha_c Re_d)}{\rho_d \alpha_c^2 D_d^2} (\mathbf{u}_{\rm c} - \mathbf{u}_{\rm d}) & \alpha_c < 0.8 \\ \frac{0.75 m_d \mu_c Re_d C_{D,correction}}{\rho_d \alpha_c^{2.65} D_d^2} (\mathbf{u}_{\rm c} - \mathbf{u}_{\rm d}) & \alpha_c \geq 0.8 \end{cases}. \tag{18}$$

The corrected drag coefficient [8]

$$C_{D,correction} = C_D(Re_d, Ma_d) \cdot C_1(\alpha_d). \tag{19}$$

Here $Ma_d \equiv |\mathbf{u}_{\rm c} - \mathbf{u}_{\rm d}|/c$ is the particle Mach number, and $c$ is the sound speed of the gas mixture. Parmar et al. [40] propose an improved correlation for the drag coefficient $C_D(Re_d, Ma_d)$, which is valid for cases with $Re_d \leq 2 \times 10^5$ and $Ma_d \leq 1.75$

$$C_D = \begin{cases} C_{D,std} + \left[C_{D,Ma_{cr}} - C_{D,std}\right] \frac{Ma_d}{0.6}, & \text{if } Ma_d \leq 0.6, \\ C_{D,sub}, & \text{if } 0.6 \leq Ma_d \leq 1.0, \\ C_{D,sup}, & \text{if } 1.0 \leq Ma_d \leq 1.75, \end{cases} \tag{20}$$

$$C_{D,std} = \frac{24}{Re_d}(1 + 0.15 Re_d^{0.687}) + 0.42\left(1 + \frac{42500}{Re_d^{1.16}}\right)^{-1}. \tag{21}$$

The expressions of $C_{D,Ma_{cr}}$, $C_{D,sub}$, and $C_{D,sup}$ can be found in Ref. [40]. Besides, the correction function $C_1$ is a function of the particle volume fraction [41]

$$C_1(\alpha_d) = \frac{1+2\alpha_d}{(1-\alpha_d)^2}. \tag{22}$$

### 2.3 Multiphase particle-in-cell method

In the Multiphase Particle-In-Cell (MP-PIC) method [29, 42], the particle phase is described by a probability distribution function, and the particle collision is modelled in the Eulerian framework. Here we only adopt the collision model from the MP-PIC method to describe particle collisions in moderately dense flows. The particle velocity is updated sequentially considering fluid dynamic force or collision force. Specifically, the particle velocity at the $(n+1)$ time step is

$$\mathbf{u}_{\rm d}^{\rm n+1} = \bar{\mathbf{u}}_{\rm d} + \mathbf{u}_{\rm p\tau}. \tag{23}$$

$\bar{\mathbf{u}}_{\rm d}$ is the intermediate velocity, calculated only with the fluid dynamic force $\mathbf{F}_{\rm surf}$, i.e.,

$$\bar{\mathbf{u}}_{\rm d} = \mathbf{u}_{\rm d}^{\rm n} + \frac{\mathbf{F}_{\rm surf}}{m_d} \cdot \Delta t, \tag{24}$$

where $\mathbf{u}_{\rm d}^{\rm n}$ is the particle velocity from previous step, and $\Delta t$ is the time step.

In Eq. (23), $\mathbf{u}_{\rm p\tau}$ is the correction velocity, solved from Eq. (13) when only the particle collision force is considered [42]. In the orthogonal directions $\mathbf{e}_{\rm k} = (\mathbf{e}_{\rm x}, \mathbf{e}_{\rm y}, \mathbf{e}_{\rm z})$, we have

$$u_{p\tau_k} = \begin{cases} \min\left(\Delta\mathbf{u}_{\rm p\tau} \cdot \mathbf{e}_{\rm k}, -(1+\gamma)(\bar{\mathbf{u}}_{\rm d} - \tilde{\mathbf{u}}_{\rm d}) \cdot \mathbf{e}_{\rm k}\right), & \text{if } \mathbf{F}_{\rm pp} \cdot \mathbf{e}_{\rm k} > 0, (\bar{\mathbf{u}}_{\rm d} - \tilde{\mathbf{u}}_{\rm d}) \cdot \mathbf{e}_{\rm k} < 0, \\ \max\left(\Delta\mathbf{u}_{\rm p\tau} \cdot \mathbf{e}_{\rm k}, -(1+\gamma)(\bar{\mathbf{u}}_{\rm d} - \tilde{\mathbf{u}}_{\rm d}) \cdot \mathbf{e}_{\rm k}\right), & \text{if } \mathbf{F}_{\rm pp} \cdot \mathbf{e}_{\rm k} < 0, (\bar{\mathbf{u}}_{\rm d} - \tilde{\mathbf{u}}_{\rm d}) \cdot \mathbf{e}_{\rm k} > 0, \\ 0, & \text{otherwise.} \end{cases} \tag{25}$$

$u_{p\tau_k}$ is the *k*-component of $\mathbf{u}_{\rm p\tau}$. $\Delta\mathbf{u}_{\rm p\tau}$ is the initial estimation of the correction velocity, and $\tilde{\mathbf{u}}_{\rm d}$ is the mass-averaged particle velocity vector in a CFD cell, from Eq. (7), and here termed as *cloud velocity*. In closely packed regions where the particle volume fraction $\alpha_d$ approaches the packing limit $\alpha_{pl}$ (0.65 for spherical particles), $\Delta\mathbf{u}_{\rm p\tau}$ will be extremely large [43]. Therefore, a limiter is applied in Eq. (25), and $\gamma$ is the restitution coefficient and takes 0.9 [42]. The physical interpretation of Eq. (25) will be detailed in Fig. 2 and

Table *1*.

Rather than updating $\bar{\mathbf{u}}_{\rm d}$ under the Lagrangian framework (i.e., Eq. 24), the MP-PIC method models the collision forces under the Eulerian framework [29, 42]. Specifically, the solid stress, $\tau_{coll}$, is computed over the Eulerian grid based on the reconstructed particle volume fraction and other properties. In this work, we follow Harris and Crighton [44], i.e.,

$$\tau_{coll} = \frac{P_s \alpha_d^{\beta}}{\max\left[\alpha_{pl} - \alpha_d,\ \varepsilon(1-\alpha_d)\right]}, \tag{26}$$

where the pressure constant $P_s$ is $8 \times 10^5$ Pa [8], and $\beta$ is 3.0 [42]. $\varepsilon = 1 \times 10^{-7}$ is used to ensure numerical stability. Therefore, the collision force $\mathbf{F}_{\rm pp}$ can be modelled as [42]

$$\mathbf{F}_{\rm pp} = -\frac{V_d}{\alpha_d} \nabla \tau_{coll}. \tag{27}$$

Based on Eq. (13) only with the collision force $\mathbf{F}_{\rm pp}$, the $\Delta\mathbf{u}_{\rm p\tau}$ in Eq. (25) is

$$\Delta\mathbf{u}_{\rm p\tau} = -\frac{\Delta t \cdot \nabla \tau_{coll}}{\rho_d \alpha_d}. \tag{28}$$

The particle collision modelling in the MP-PIC method can be interpreted with Fig. 2. For simplicity, we only consider the $x$-direction and extension to multi-dimensional problems is straightforward. We assume that the right direction is positive. **C** denotes the particle cloud, and its properties (e.g., $x$-component cloud speed $\tilde{u}_d$) are estimated based on the particles in this cell and stored at the cell center. There are four particles 1-4 in the cell, with the $x$-component velocities $\bar{u}_{di}$ ($i$ = 1, …, 4). In the MP-PIC method, $\alpha_d$ is a Eulerian quantity at the cell centre and needs to be interpolated to the particle position to model the collision force. Moreover, $\alpha_d$ at the centre **C** would be larger than that on both sides in a CFD cell, leading to the opposite sign of $\nabla\alpha_d$ at different sides. Note that the sign of $\mathbf{F}_{\mathrm{pp}}$ is always opposite to that of $\nabla\alpha_d$, according to Eqs. (26) and (27). In this sense, the sign of $\nabla\alpha_d$ is consistent with physical rationales, i.e., $\mathbf{F}_{\mathrm{pp}}$ is opposite in different sides. Specifically, for the particles left to the cell centre (e.g., particles 1 and 3 in Fig. 2), $\mathbf{F}_{\mathrm{pp}}$ points leftward. For the particles right to the cell centre (e.g., particles 2 and 4), $\mathbf{F}_{\mathrm{pp}}$ points rightward. In Eq. (25), if $\mathbf{F}_{\mathrm{pp}} \cdot \mathbf{e}_{\mathrm{k}} > 0$ and $(\bar{\mathbf{u}}_{\mathrm{d}} - \tilde{\mathbf{u}}_{\mathrm{d}}) \cdot \mathbf{e}_{\mathrm{k}} < 0$, it indicates $(\bar{\mathbf{u}}_{\mathrm{d}} - \tilde{\mathbf{u}}_{\mathrm{d}})$ and $\mathbf{F}_{\mathrm{pp}}$ are in the opposite direction, then the particle is approaching the cloud and particle collision occurs. Also, the particle collision would happen when $\mathbf{F}_{\mathrm{pp}} \cdot \mathbf{e}_{\mathrm{k}} < 0$ and $(\bar{\mathbf{u}}_{\mathrm{d}} - \tilde{\mathbf{u}}_{\mathrm{d}}) \cdot \mathbf{e}_{\mathrm{k}} > 0$.

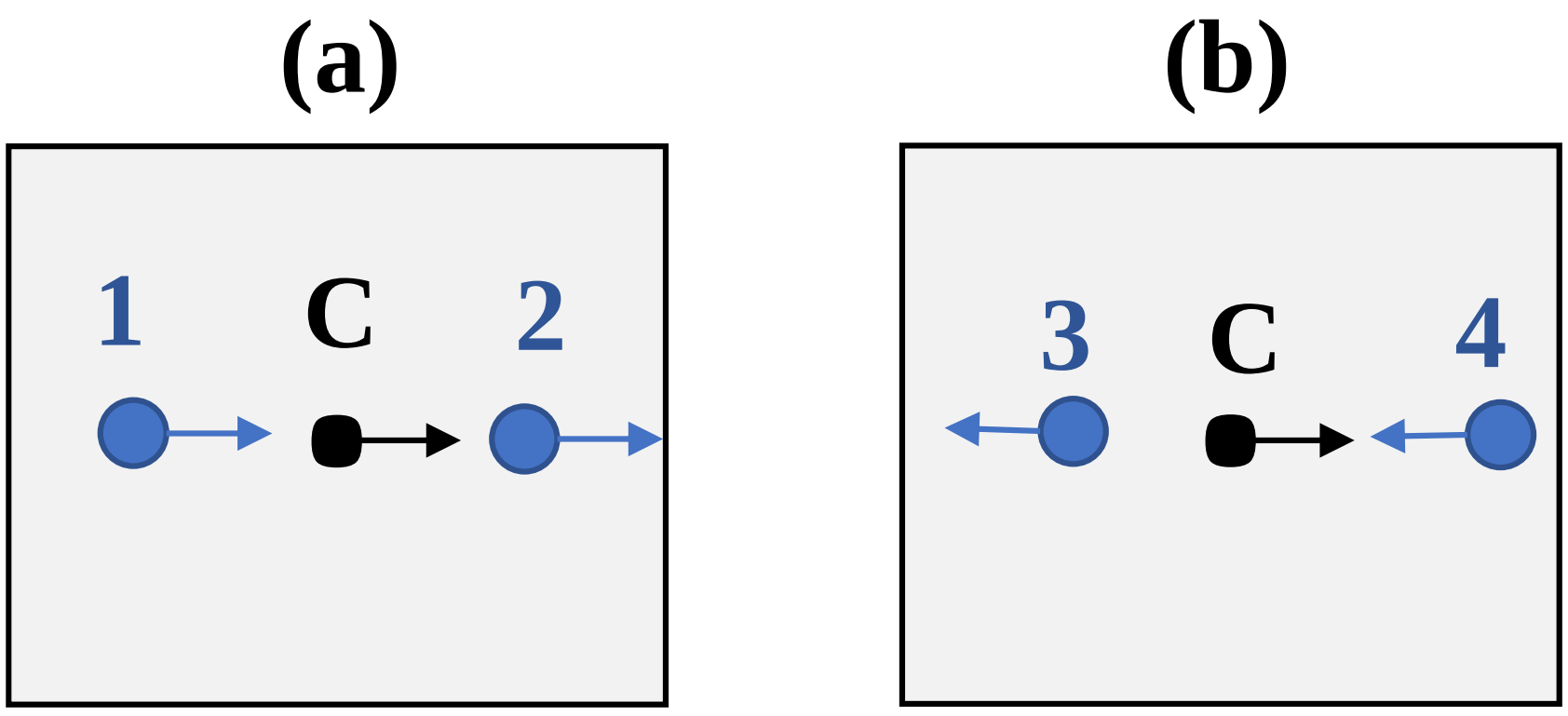

Fig. 2 A schematic showing the MP-PIC model. 1-4: particles, **C**: particle cloud.

If the particle cloud velocity $\tilde{\mathbf{u}}_{\mathrm{d}}$ is assumed to be positive, as shown in Fig. 2, there are six cases, which are listed in Table *1*. $u_{p\tau} = 0$ means that particle collision does not occur. Note that when $\tilde{\mathbf{u}}_{\mathrm{d}}$ is negative, similar six cases can also be obtained according to Eq. (25).

Table 1. Particle collision situations in the MP-PIC method.

| Case | Description | Relative motion of particle and cloud | Outcome | Condition |
|---|---|---|---|---|
| 1 | $0 < \bar{u}_{d1} \leq \tilde{u}_d$ | leaving | $u_{p\tau} = 0$ | $\mathbf{F}_{\mathrm{pp}} \cdot \mathbf{e}_{\mathrm{k}} < 0, (\bar{\mathbf{u}}_{\mathrm{d1}} - \tilde{\mathbf{u}}_{\mathrm{d}}) \cdot \mathbf{e}_{\mathrm{k}} \leq 0$ |
| 2 | $\bar{u}_{d1} > \tilde{u}_d > 0$ | approaching | $u_{p\tau} < 0$ | $\mathbf{F}_{\mathrm{pp}} \cdot \mathbf{e}_{\mathrm{k}} < 0, (\bar{\mathbf{u}}_{\mathrm{d1}} - \tilde{\mathbf{u}}_{\mathrm{d}}) \cdot \mathbf{e}_{\mathrm{k}} > 0$ |
| 3 | $0 < \bar{u}_{d2} < \tilde{u}_d$ | approaching | $u_{p\tau} > 0$ | $\mathbf{F}_{\mathrm{pp}} \cdot \mathbf{e}_{\mathrm{k}} > 0, (\bar{\mathbf{u}}_{\mathrm{d2}} - \tilde{\mathbf{u}}_{\mathrm{d}}) \cdot \mathbf{e}_{\mathrm{k}} < 0$ |
| 4 | $\bar{u}_{d2} \geq \tilde{u}_d > 0$ | leaving | $u_{p\tau} = 0$ | $\mathbf{F}_{\mathrm{pp}} \cdot \mathbf{e}_{\mathrm{k}} > 0, (\bar{\mathbf{u}}_{\mathrm{d2}} - \tilde{\mathbf{u}}_{\mathrm{d}}) \cdot \mathbf{e}_{\mathrm{k}} \geq 0$ |
| 5 | $\bar{u}_{d3} < 0, \tilde{u}_d > 0$ | leaving | $u_{p\tau} = 0$ | $\mathbf{F}_{\mathrm{pp}} \cdot \mathbf{e}_{\mathrm{k}} < 0, (\bar{\mathbf{u}}_{\mathrm{d3}} - \tilde{\mathbf{u}}_{\mathrm{d}}) \cdot \mathbf{e}_{\mathrm{k}} < 0$ |
| 6 | $\bar{u}_{d4} < 0, \tilde{u}_d > 0$ | approaching | $u_{p\tau} > 0$ | $\mathbf{F}_{\mathrm{pp}} \cdot \mathbf{e}_{\mathrm{k}} > 0, (\bar{\mathbf{u}}_{\mathrm{d4}} - \tilde{\mathbf{u}}_{\mathrm{d}}) \cdot \mathbf{e}_{\mathrm{k}} < 0$ |

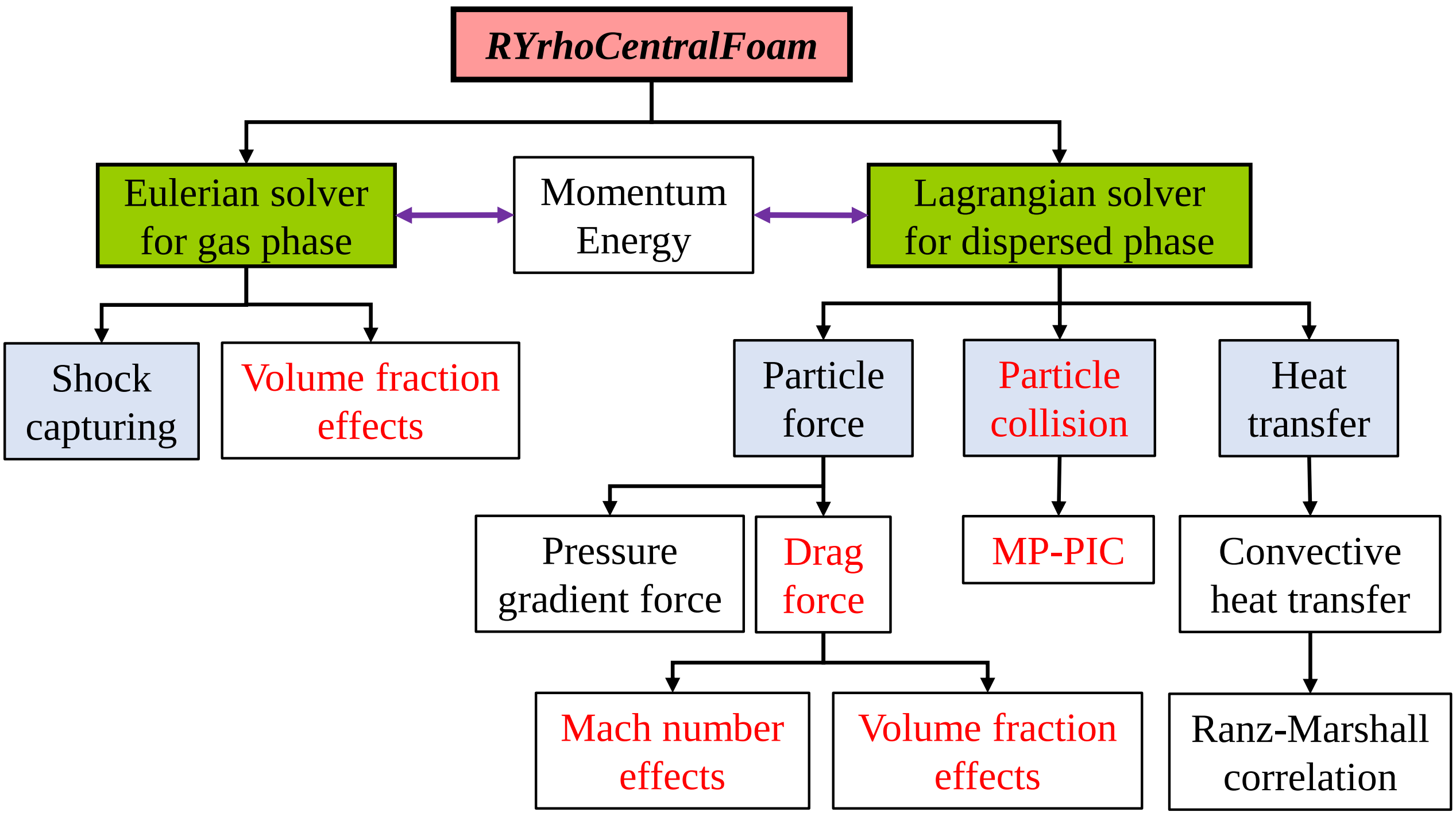

Fig. 3 Key modules of the *RYrhoCentralFoam* solver. Red blocks are new developments for moderately dense gas-particle two-phase flows.

## 3 NUMERICAL IMPLEMENTATIONS

The governing equations of the gas and dispersed phases are solved with a compressible flow solver *RYrhoCentralFoam*, developed from OpenFOAM 6.0. Compared to our previous solver for dilute two-phase flows [2, 23, 45-47], new developments for moderately dense flows are included, i.e., volume fraction effects in the gas phase equations, improved drag model (particle Mach number and volume fraction effects), and MP-PIC collision model. The key modules of the new *RYrhoCentralFoam* solver are shown in Fig. 3. Coupling of the Eulerian and Lagrangian solvers is realized with runtime exchanges of momentum and energy.

The gas phase equations, i.e., Eqs. (2)-(4), are discretized with the finite-volume method. The second-order backward scheme is used for temporal discretization, and the time step is about $10^{-8}$ s. The second-order central differencing is applied for the diffusive fluxes. Flow discontinuities are captured with a central-upwind scheme [48] and van Leer limiter. The particle equations, i.e., Eqs. (13) and (14), are solved with the first-order Euler implicit method. The source terms in the gas phase equations, i.e., Eqs. (3) and (4), are integrated with a semi-implicit method.

## 4 SOLVER VALIDATION

*RYrhoCentralFoam* has been validated against theoretical or experimental data for purely gaseous and dilute two-phase compressible reactive flows [23, 24, 49]. In this section, we further validate the new implementations for moderately dense two-phase flows as illustrated in Fig. 3.

### 4.4.1 Ling et al. experiment

We first validate the *RYrhoCentralFoam* solver for shock-particle interactions against the experiment by Ling et al. [8]. A Mach 1.66 shock (shock speed $\mathbf{U}_s$ = 573 m/s) propagates in the 1D domain and enters a curtain of soda lime particles. The initial pressure and temperature of the driven gas are $p_0$ = 82,700 Pa and $T_0$ = 296.4 K, respectively. For particles, the material density is $\rho_d$ = 2.42 g/cm$^3$, the heat capacity $c_{p,d}$ = 840 J/(kg K), and the diameter $D_d$ = 115 μm. They are uniformly distributed in a 2 mm curtain with initial volume fraction of 21%.

Figures 4(a) and 4(b) respectively show that the solver can accurately reproduce the transmitted and reflected shocks and the pressure evolutions at two locations of $x_1$ = -68.6 mm and $x_2$ = 64.1 mm. We introduce the term "curtain front" in this manuscript, which is a loose term since the curtain edges may be diffuse after shock passage based on the experiments [5, 10, 13]. In this work, the upstream curtain front (UCF) and downstream curtain front (DCF) are identified from the locations of the leftmost and rightmost particles, respectively, similar to the procedures in the measurements [5, 11]. From Fig. 4(a), the predicted trajectories of both UCF and DCF are in good agreement with the measured data.

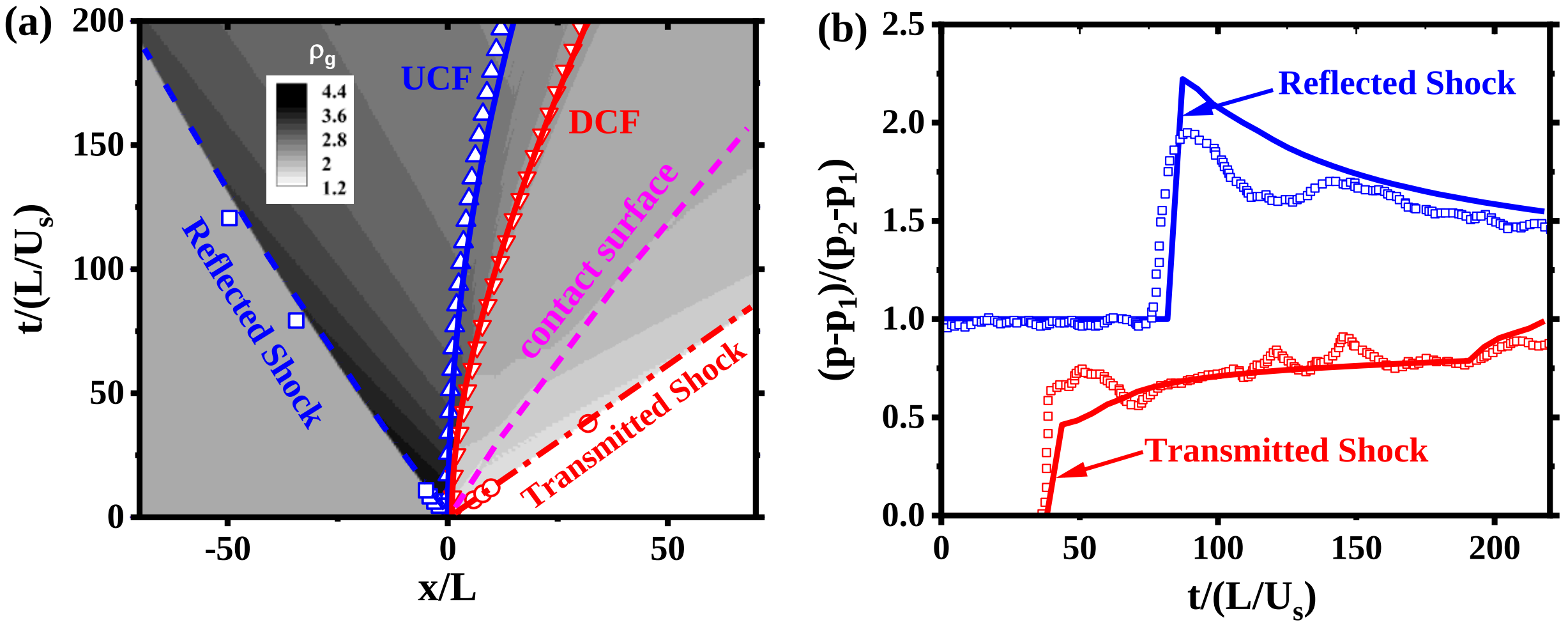

Fig. 4 Comparison of experimental and numerical results for $Ma$ = 1.66 incident shock interacting with a 2 mm particle curtain: (a) normalized $x$-$t$ diagram and (b) normalized pressure-time diagram at $x_1$ = -68.6 mm and $x_2$ = 64.1 mm. L: curtain initial width, **U**$_s$: incident shock velocity, $p_1$ and $p_2$: undisturbed pressure at $x_1$ and $x_2$, respectively. Symbols: experimental data [8], lines: *RYrhoCentralFoam*.

4.4.2 Park et al. experiment

The second validation is based on the experiment by Park et al. [50]. Different from the first case, a curtain with non-uniform particle distribution is considered, characterized by spatially varying particle volume fractions. In the simulation, the volume fraction profile is approximated by 8 bins (500 μm/bin) with particle volume fractions of 5%, 13%, 20% and 23%, as shown in Fig. 5(a). A Mach 1.67 incident shock propagates towards the right side of the shock tube. In the driven gas, the initial pressure and temperature are $p_0$ = 82700 Pa and $T_0$ = 296.4 K, respectively. The particle material density, heat capacity, and diameter are $\rho_d$ = 2.42 g/cm$^3$, $C_{p,d}$ = 840 J/(kg K), and $D_d$ = 115 μm, respectively. Fig.

5(b) shows that our solver can reasonably reproduce the movement of the upstream and downstream fronts of the curtain.

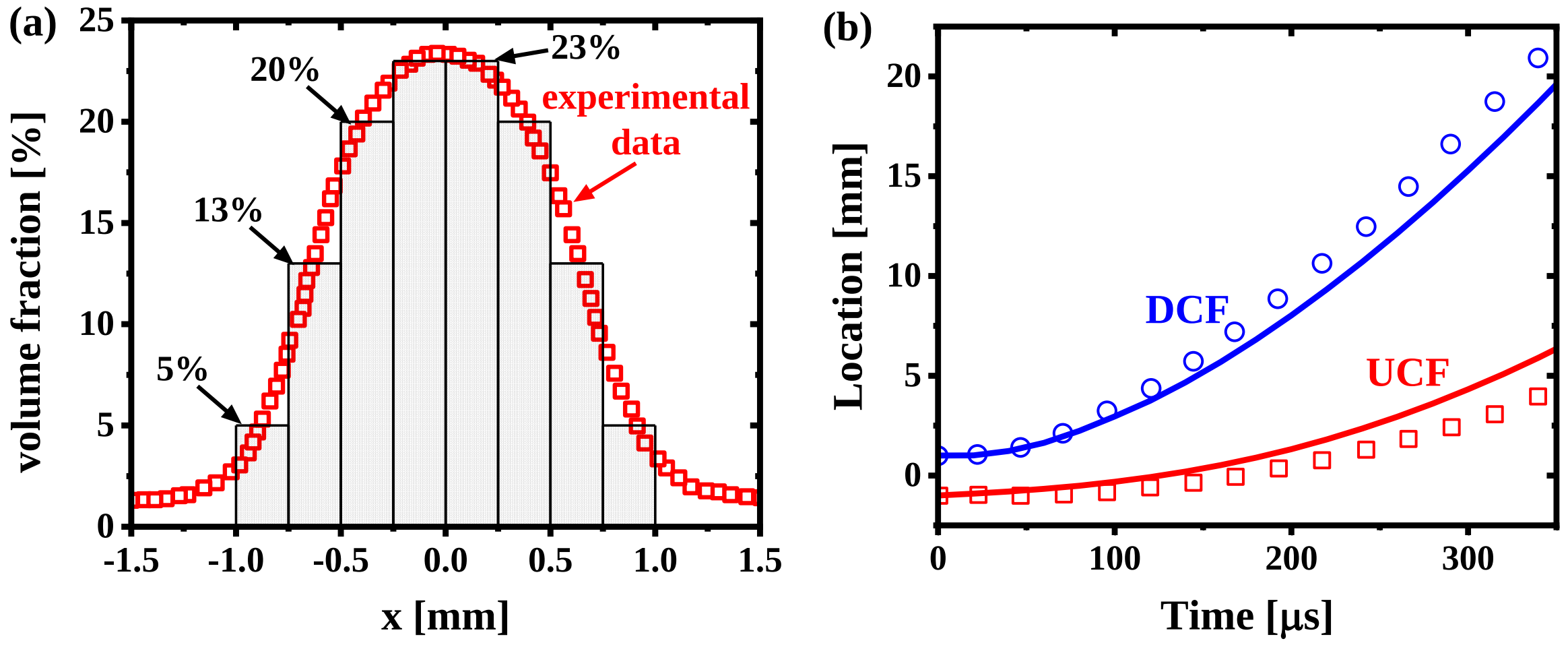

Fig. 5 (a) Particle initial volume fraction distributions; (b) Trajectories of UCF and DCF. Symbols: experimental data [50], lines: *RYrhoCentralFoam*.

## 5. PHYSICAL PROBLEM

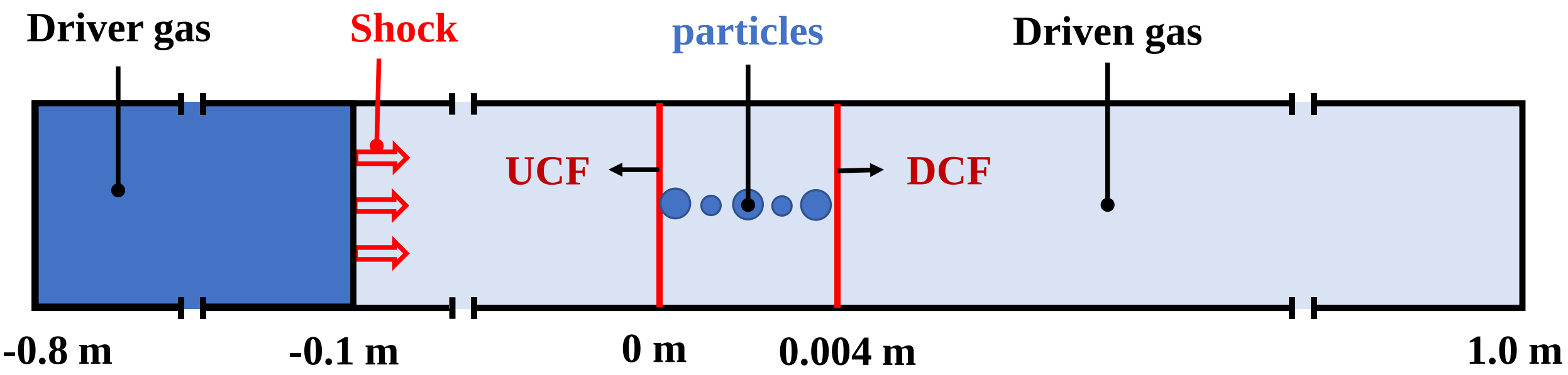

Fig. 6 Schematic of one-dimensional domain for shock-curtain interactions.

To quantitively investigate the effects of compressible gas-particle interactions and particle-particle collisions under moderately dense flow regime, a one-dimensional numerical configuration is considered (see Fig. 6). The domain is 1.8 m long and initially filled with air. For the driver gas, an incident shock wave with $Ma$ = 1.66 is initialized at $x$ = -0.1 m and transmits towards the right side. For the driven gas, the initial pressure and temperature are $p_0$ = 0.1 MPa and $T_0$ = 300 K, respectively. Zero-gradient condition is applied at the left and right boundaries.

To study the shock-curtain interactions, we vary the properties of the particle curtain, including initial volume fraction and particle size. In our following analysis, $t$ = 0 corresponds to the instant when the incident shock arrives at the UCF. As the baseline parameters, the particle material density is $\rho_d$ = 5,000 kg/m$^3$, heat capacity $c_{p,d}$ = 840 J/(kg K), and diameter $D_d$ = 200 µm. The initial thickness of the curtain is $L$ = 4 mm (i.e., the distance between UCF and DCF, see Fig. 6) and the volume fraction is $\alpha_d$ = 10%. The particles remain stationary before shock passage.

The domain is discretized with Cartesian cells, and particles are uniformly distributed across the curtain. The influences of mesh size ($\Delta x$) and particle number density ($m = N_i/\Delta x$) on the results are respectively analysed in Sections A and B of the supplementary document respectively. The results show that the mesh size makes effect on the displacement of UCF and DCF, but almost does not influence on the gas quantities (such as $\mathbf{u}_\mathrm{c}$ and $p$). For the number density effects, small $m$ results in oscillations of interphase coupling terms, e.g., $\mathbf{S}_\mathrm{mom}$. In the following, the mesh size $\Delta x$ = 500 µm and particle number density $m$ = 1×10$^5$/m will be adopted.

## 6. RESULTS AND DISCUSSION

### 6.1 Mono-dispersed particle curtain

6.1.1 Initial particle volume fraction effects

In this section, three initial particle volume fractions are considered, i.e., $\alpha_d$ = 2%, 10% and 20%. The particle diameter is $D_d$ = 200 µm, and the incident shock Mach number is $Ma$ = 1.66. The trajectories of UCF and DCF after shock impacting are shown in Fig. 7(a), whilst the fluid dynamic and collision forces on the UCF and DCF particles are demonstrated in Fig. 7(b).

For $\alpha_d$ = 2%, the DCF particle has the almost same speed as the UCF particle, because the forces exerted on two particles are close, see Fig. 7(b). As $\alpha_d$ increases, the UCF slows down, whereas the DCF trajectories are less affected. This is because the total force on the former decreases with $\alpha_d$, but stay around the same for the latter. As shown in Fig. 7(b), for the UCF particles, the fluid dynamic force $\mathbf{F}_\mathrm{surf}$ decreases after the shock passes the curtain, whereas the collision force $\mathbf{F}_\mathrm{pp}$ is always close to zero. This is reasonable, because the UCF particles always lie leftmost, and the velocity is smaller than the $\widetilde{\mathbf{u}}_\mathrm{d}$ of the local cell, corresponding to case 1 in Table *1*. Thus, no collision happens between the UCF particle and the rest. Nonetheless, for the DCF particles, $\mathbf{F}_{\mathrm{surf}}$ fluctuates strongly, particularly for large $\alpha_d$, e.g., 10% and 20%. Under denser conditions, the rightmost particle (i.e., at the DCF) may correspond to various particles at the leading front at different instants, leading to fluctuating $\mathbf{F}_{\mathrm{surf}}$. Moreover, collisions happen at some instants for $\alpha_d$ = 10% and 20%, but the $\mathbf{F}_{\mathrm{pp}}$ is generally smaller than the $\mathbf{F}_{\mathrm{surf}}$. Numerical experiments with the collision model turned off are also conducted and the analysis is presented in Section C of the supplementary documentary. It shows that the particle collision effects have almost negligible effects on the trajectories of the mono-sized particle curtains. Only observable difference is the DCF behaviors in high $\alpha_d$ cases, e.g., 10% and 20%.

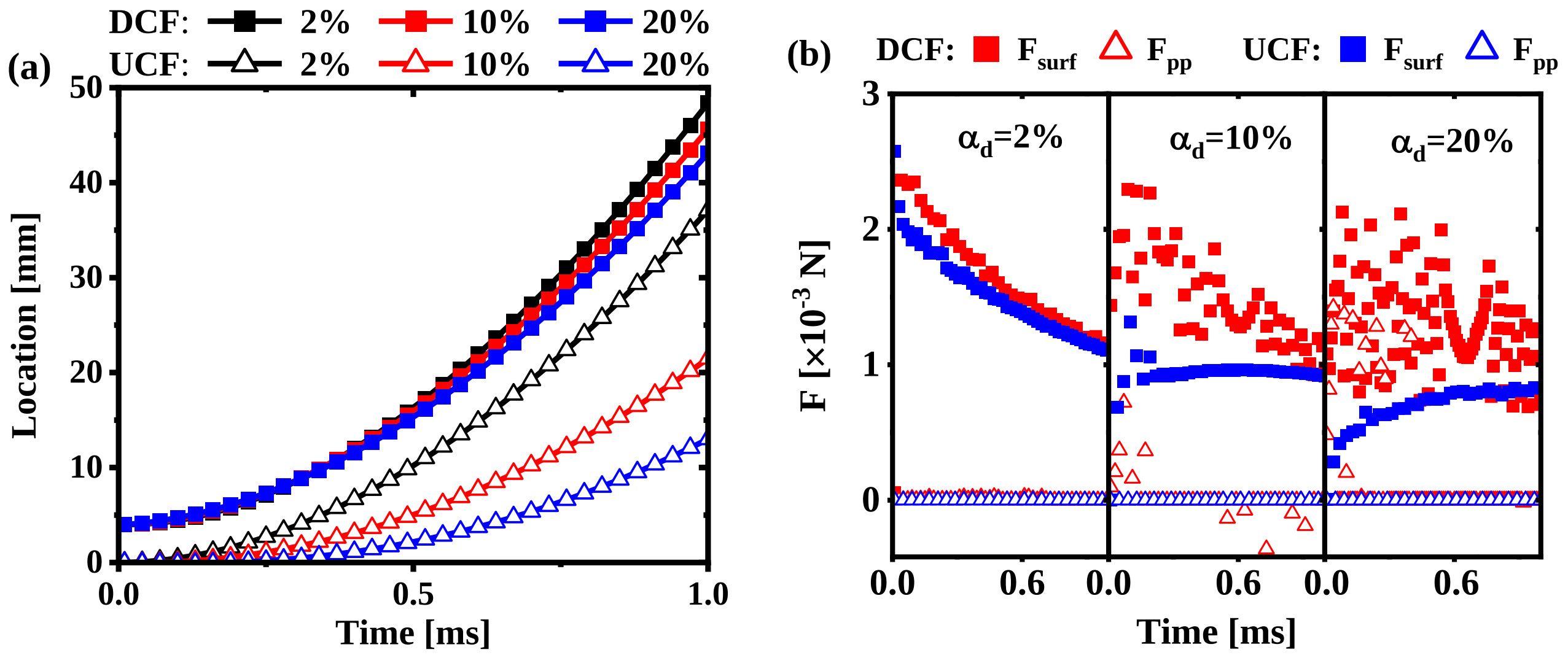

Fig. 7 (a) UCF and DCF trajectories and (b) time evolutions of UCF and DCF particle forces with different particle volume fractions.

The dimensionless time evolutions of the curtain thickness are shown in Fig. 8(a). Following Theofanous et al. [10] and DeMauro et al. [14], the dimensionless time is defined as $t^* = \mathbf{U}_{\Delta\mathrm{p}} t \alpha_d^{0.25}/L_0$ with $L_0$ being the initial particle curtain thickness. The pressure-based velocity $\mathbf{U}_{\Delta\mathrm{p}}$ is around 11.75 m/s in the current case. The simulation time of three cases is 2 ms. The shock leaves the curtain at $t^*$~0.02. As shown in the inset of Fig. 8(a), the curtain remains stationary (i.e., both fronts are intact) at the first stage, i.e., $0 < t^* < 0.2$. It is termed as constant thickness regime (CTR), also observed by DeMauro et al [14]. The time lag between shock wave arrival and initial curtain motion arises from the inertia of relatively coarse particles (200 μm). After the CTR, the curtain begins to expand, and the expansion velocity ($\mathbf{U}_\mathrm{L}$) is shown in Fig. 8(b). At this stage ($0.2 < t^* < 2.0$), the curtain thickness increases faster as the $\mathbf{U}_\mathrm{L}$ becomes larger. However, this stage is a varying-acceleration stage, which can be seen the evolutions of the acceleration $\mathbf{a}_\mathrm{L}$ in Fig. 8(c). When the initial $\alpha_d$ increases to 10% and 20%, the curtain thickness increases faster. This is mainly caused by the slowdown of the UCF particle, see Fig. 7(a). At the third stage ($2.0 < t^* < 3.0$), the expansion velocity $\mathbf{U}_\mathrm{L}$ increases to the largest value and the expansion acceleration $\mathbf{a}_\mathrm{L}$ drops to zero. This stage is termed as constant-velocity regime (CVR), also observed in the experiments [10, 14]. However, when $t^* > 3.0$, the $\mathbf{a}_\mathrm{L}$ of $\alpha_d = 20\%$ gradually decrease to below zero, leading to a reduced expansion velocity of the particle curtain.

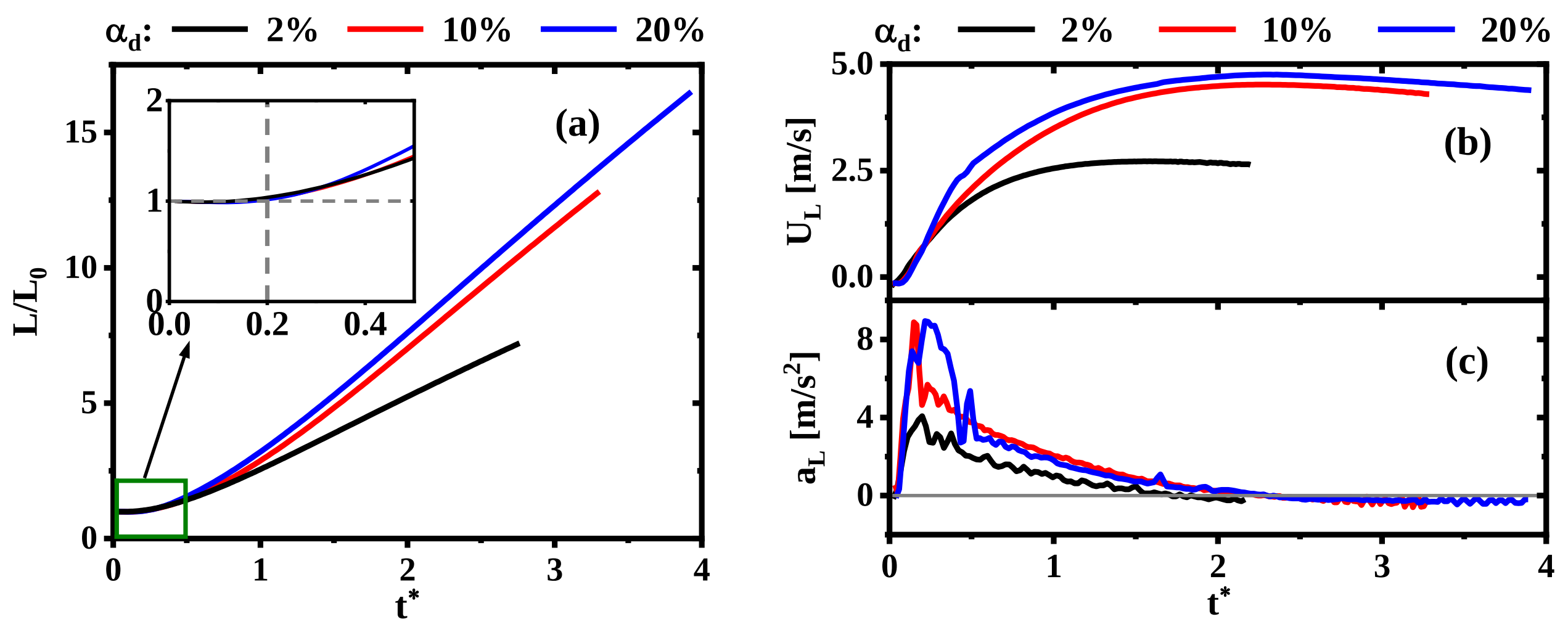

Fig. 8 Time evolutions of (a) normalized curtain thickness, (b) expansion velocity, and (c) expansion acceleration with different particle volume fractions.

The aerodynamic behaviors along the two-phase interface largely depend on the difference of the thermal properties between two media [51]. With the assumptions of zero particle Stokes number and no particle effects on the gas pressure, the gas-particle mixture in the curtain can be deemed "equivalent gas". As such, the effective density, heat capacity ratio, gas constant, sound speed, and acoustic impedance of the two-phase mixture can be estimated from [8, 52]

$$\rho_e = \alpha_c \rho_c + \alpha_d \rho_d, \tag{29}$$

$$\gamma_e = \gamma \frac{1+\frac{\alpha_d \rho_d c_{p,d}}{\alpha_c \rho_c \, c_{p,c}}}{1+\gamma \frac{\alpha_d \rho_d c_{p,d}}{\alpha_c \rho_c \, c_{p,c}}}, \tag{30}$$

$$R_e = \frac{\alpha_c \rho_c}{\alpha_c \rho_c + \alpha_d \rho_d} R, \tag{31}$$

$$c_e = \sqrt{\gamma_e R_e T}, \tag{32}$$

$$Z_e = \rho_e c_e, \tag{33}$$

where the subscript $e$ denotes equivalent gas-particle mixture. $c_{p,c}$ is the specific heat capacity of the background gas at constant pressure, and $\gamma$ is the heat capacity ratio of the gas phase, taking 1.4. The acoustic impedance ($Z$) in Eq. (33) is a measure of material acoustic stiffness. The change in acoustic impedance between media 1 and 2 at the interface is impedance mismatch, i.e., $\delta Z = Z_2 - Z_1$. If a shock transmits from media 1 to 2 with $\delta Z > 0$, the reflected wave will be a shock wave. A larger $\delta Z$ leads to a stronger shock reflection. Otherwise, the reflected wave is an expansion fan [51, 53].

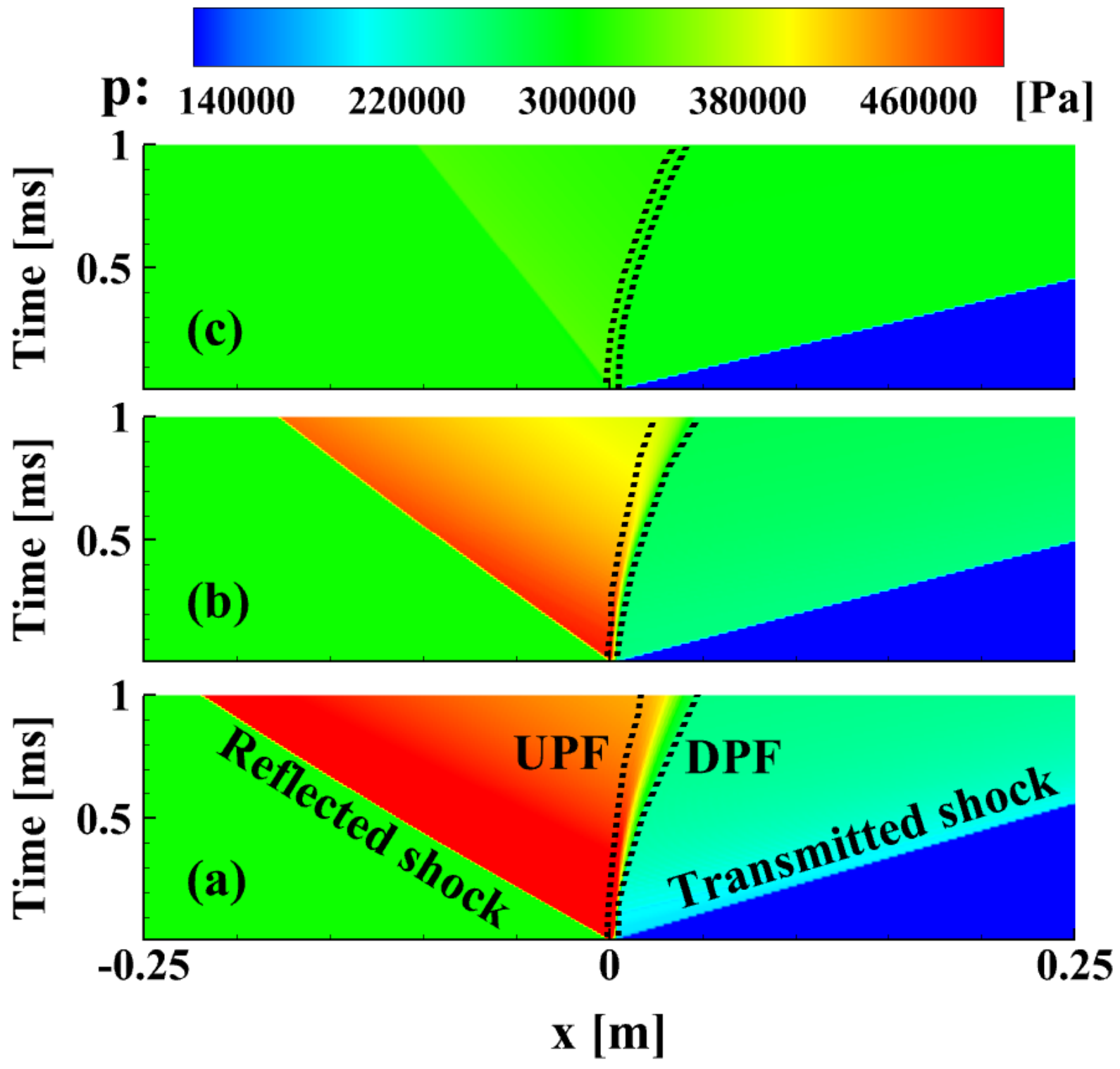

Fig. 9 $x$-$t$ diagram of gas pressure with various particle volume fractions: (a) 20%, (b) 10%, and (c) 2%.

Figure 9 shows the $x$-$t$ diagrams of gas pressure evolutions when different particle volume fractions are considered, and Fig. 10 plots the spatial profiles of gas velocity and pressure at one typical instant. We take $\alpha_d$ = 20% as an example. In this case, the acoustic impedance of the gas-particle mixture, i.e., particle curtain region, is $Z_e$ = 8,950 kg/(m$^2$ s), whereas that of the surrounding gas is $Z_c$ = 403.3 kg/(m$^2$ s). As shown in Fig. 9, when the shock enters the curtain at the UCF ($\delta Z > 0$), part of the shock reflects off the UCF, whilst part of it transmits into the curtain. When the transmitted shock leaves the curtain, the impedance difference across the DCF is $\delta Z < 0$. Therefore, an expansion fan is generated towards the curtain, which can be seen from the positive velocity gradient and negative pressure gradient across the curtain in Fig. 10. This leads to curtain expansion due to the gas velocities and pressure difference between the UCF and DCF. This is also observed by DeMauro et al. [13] and Theofanous et al. [10] through the experimental measurements.

When $\alpha_d$ decreases, the degree of the acoustic mismatch at both UCF and DCF decreases. Therefore, at the UCF, the reflected shock is accordingly weakened. For instance, when $\alpha_d$ = 2%, the reflected shock is marginal, because the impedance difference is the smallest among these cases, i.e., $\delta Z$ = 2,745 kg/(m$^2$ s). For the transmitted shock, decreasing $\alpha_d$ leads to higher intensity and hence larger wave speed. This is seen from the shock pressure and location in Fig. 10. At the DCF, the expansion fan becomes weak when $\alpha_d$ is low, which can be confirmed by the reduced magnitudes of pressure and velocity gradient near the DCF, as revealed in Fig. 10. This weakened expansion fan accounts for the reduced curtain expansion with small $\alpha_d$ (see Fig. 7a).

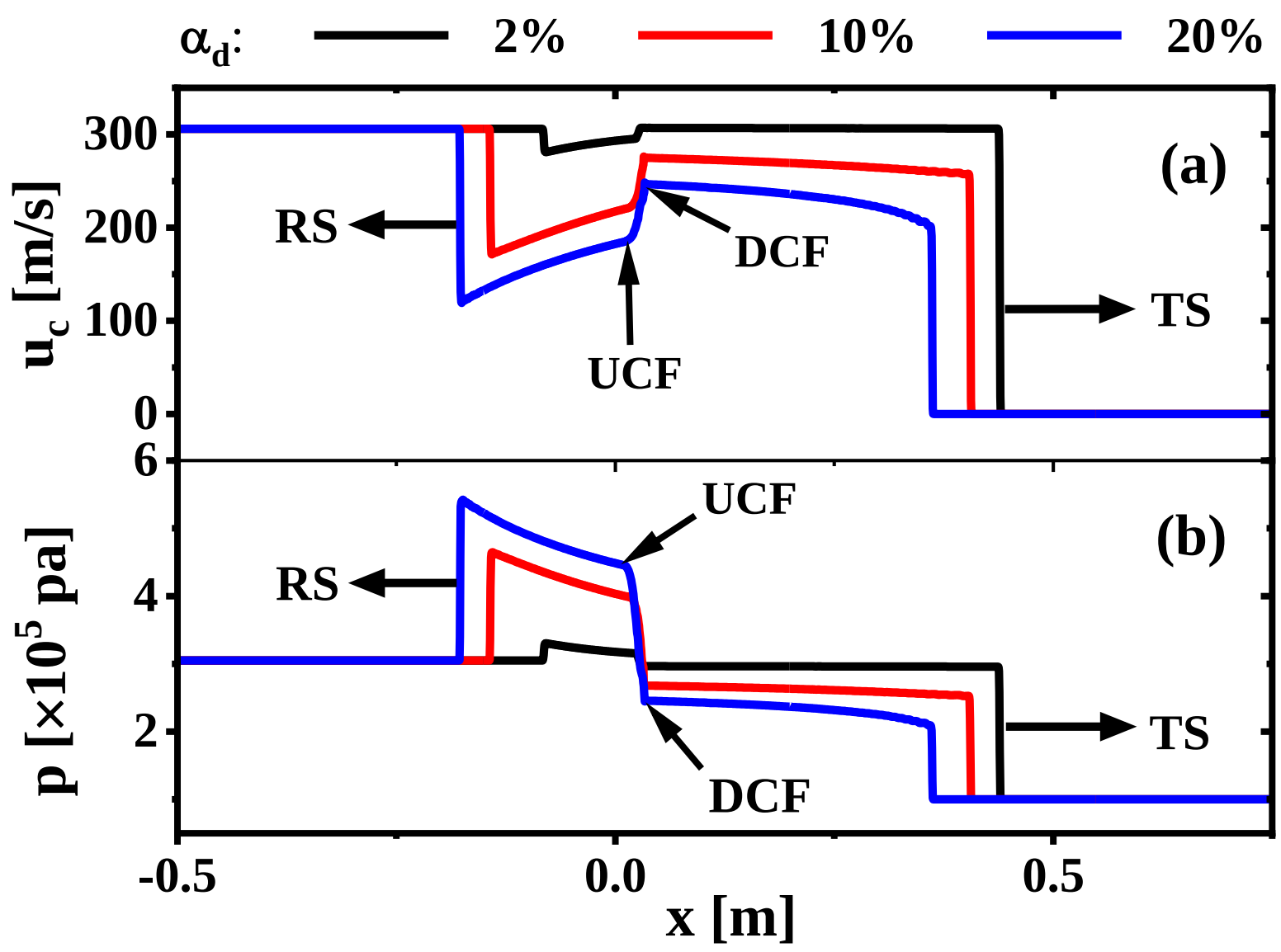

Fig. 10 Distributions of (a) gas velocity and (b) pressure with different initial particle volume fractions. RS: reflected shock; TS: transmitted shock. Data from 800 μs in Fig. 9.

### 6.1.2 Particle size effects

To study the particle size effects, three different particle diameters are considered, i.e., 50, 200, and 400 μm. The initial particle volume fraction is $\alpha_d$ = 10%. Figures 11(a) and 11(b) respectively show the trajectories and forces of the UCF and DCF particles in three cases. In all cases, $\mathbf{F}_{\mathrm{pp}}$ of the DCF particles fluctuates around zero, and generally much smaller than $\mathbf{F}_{\mathrm{surf}}$. This can be attributed to the same reason discussed in section 6.1.1. For $D_d$ = 50 μm, the total force, i.e., $\mathbf{F}_{\mathrm{surf}}$ and $\mathbf{F}_{\mathrm{pp}}$, on the DCF is larger than that on the UCF, and thus the former moves faster, see Fig. 11(a). Moreover, both UCF and DCF move fastest (slowest) for the smallest (largest) particles. Such differences can be explained from Fig. 11(b): as the particle diameter increases, although the sum of $\mathbf{F}_{\mathrm{surf}}$ and $\mathbf{F}_{\mathrm{pp}}$ on the DCF and UCF increases, the resulting particle acceleration generally decreases. This is because the mass increases faster (due to large size) than the forces with $D_d$, leading to a decreasing acceleration. The collision effects on the curtain fronts are further analyzed when the collision model is excluded, as shown in section C of the supplementary documentary. The results show that the collision model only influences the 50 μm DCF, whereas for larger particles (e.g., 200 and 400 μm) particle collision effects are negligible.

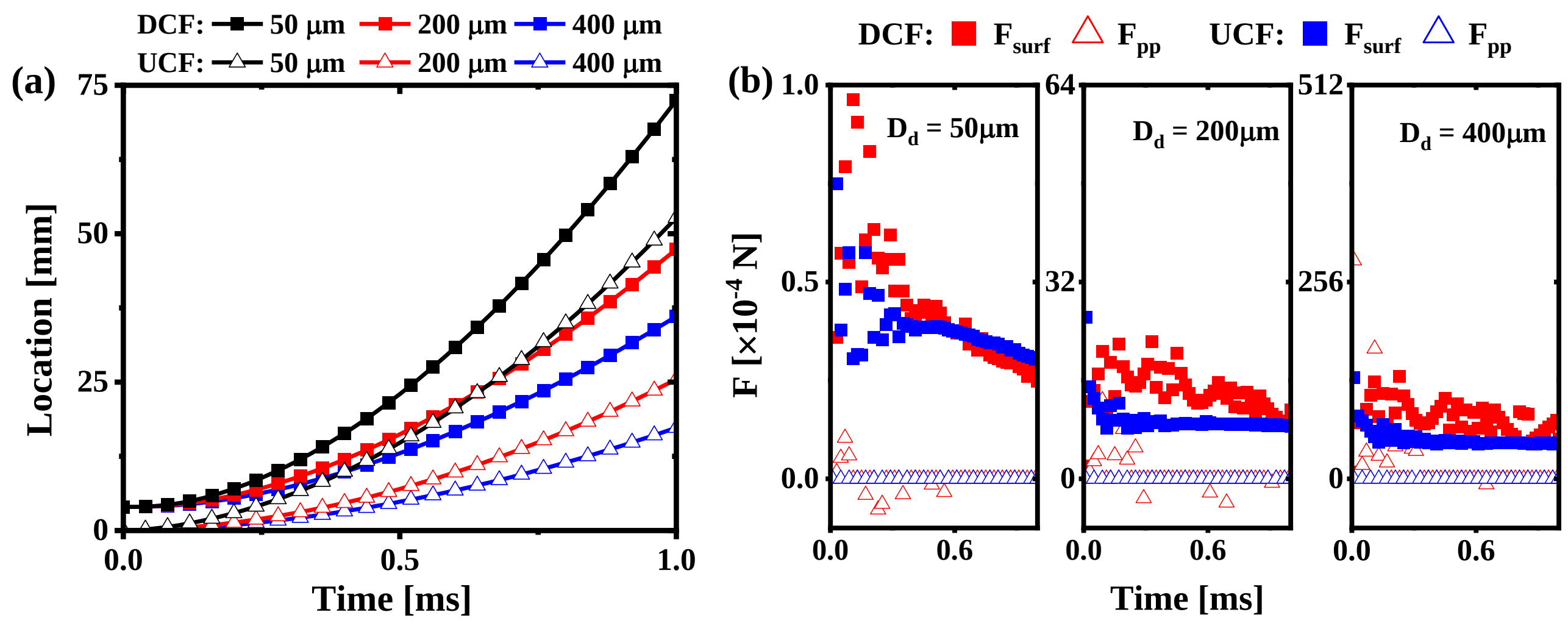

Fig. 11 (a) UCF and DCF trajectories and (b) UCF and DCF particle forces with different particle diameters.

Time evolutions of the curtain thickness, expansion velocity and acceleration in the above cases are shown in Fig. 12(a)-12(c), respectively. The shock leaves the curtain at $t^* = 0.02$. At the first stage ($0 < t^* < 0.2$), the 200 and 400 μm particle curtains remain stationary, but the 50 μm one exhibits weak compression (i.e., $L/L_0 < 1$), see the inset of Fig. 12(a). Such difference is caused by the faster momentum response time ($\tau_v$) of the smaller particles at UCF, which can be estimated from $\tau_v = (\rho_d D_d^2)/(18\mu_c)$. At the initial stage, the expansion acceleration $\mathbf{a}_\mathrm{L}$ and velocity $\mathbf{U}_\mathrm{L}$ of the 50 μm particle curtain are negative, see Fig. 12(c) and 12(b), leading to the foregoing curtain compression. The force on the particles is shown in Section D of supplementary document, and it is found that the UCF particles have larger $\mathbf{F}_\mathrm{surf}$ and thus faster motion. Note that the particle diameter is not involved in the time scaling for $t^*$. Therefore, the particle size effects on the curtain front trajectories are not considered by Theofanous et al. [10] and DeMauro et al. [14]. Furthermore, the particles in their experiments are relatively coarse (100-1000 μm), and therefore the initial curtain compression is not observed.

After around $t^* = 0.2$, three curtains begin to expand, and the expansion velocities gradually increase; see $\mathbf{U}_\mathrm{L}$ in Fig. 12(b). At the initial stage, the acceleration $\mathbf{a}_\mathrm{L}$ increases sharply. The $\mathbf{U}_\mathrm{L}$ of the 50 μm particle curtain firstly peaks at $t^* = 1.5$, and then decreases. It takes longer time for the coarser particle curtains to reach the largest $\mathbf{U}_\mathrm{L}$: it is $t^* = 2.0$ for the 200 μm curtain, while $t^* = 3.0$ for the 400 μm one.

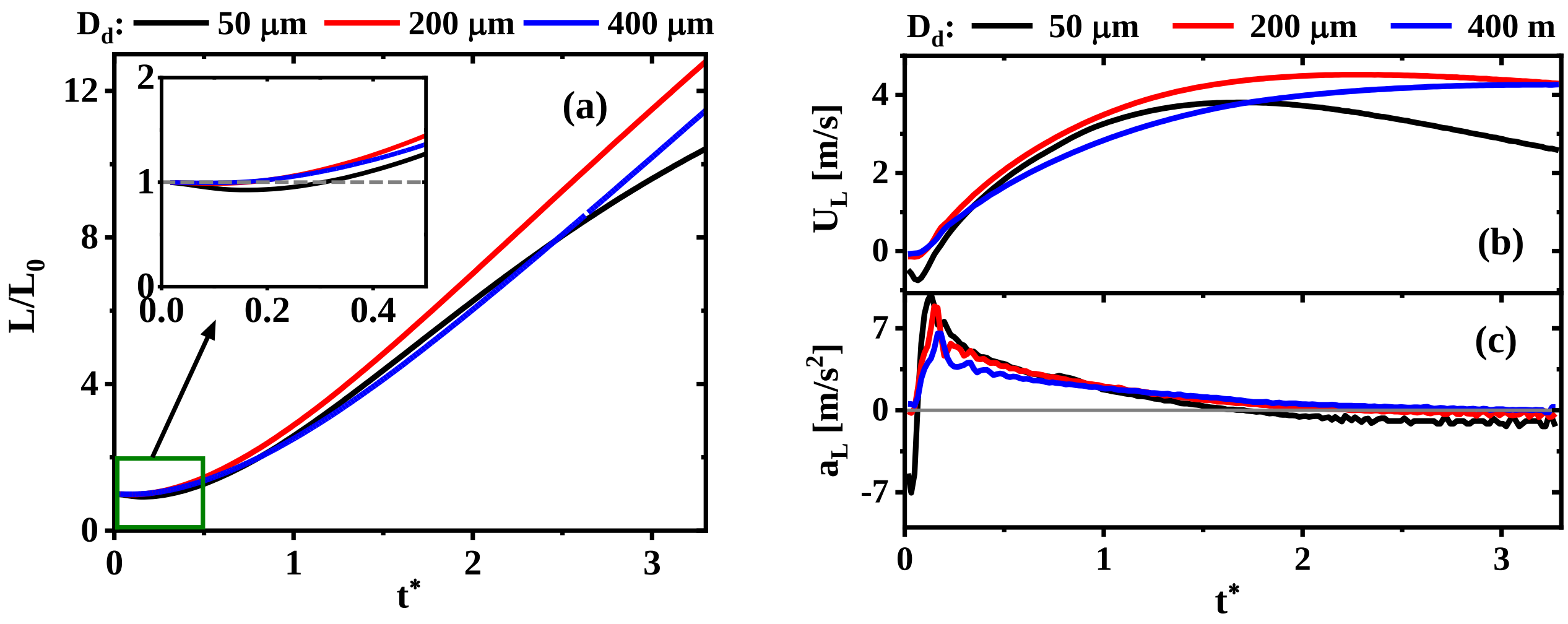

Fig. 12 Time evolutions of (a) curtain thickness, (b) expansion velocity, and (c) expansion acceleration with different particle diameters. $\alpha_d$ = 10%.

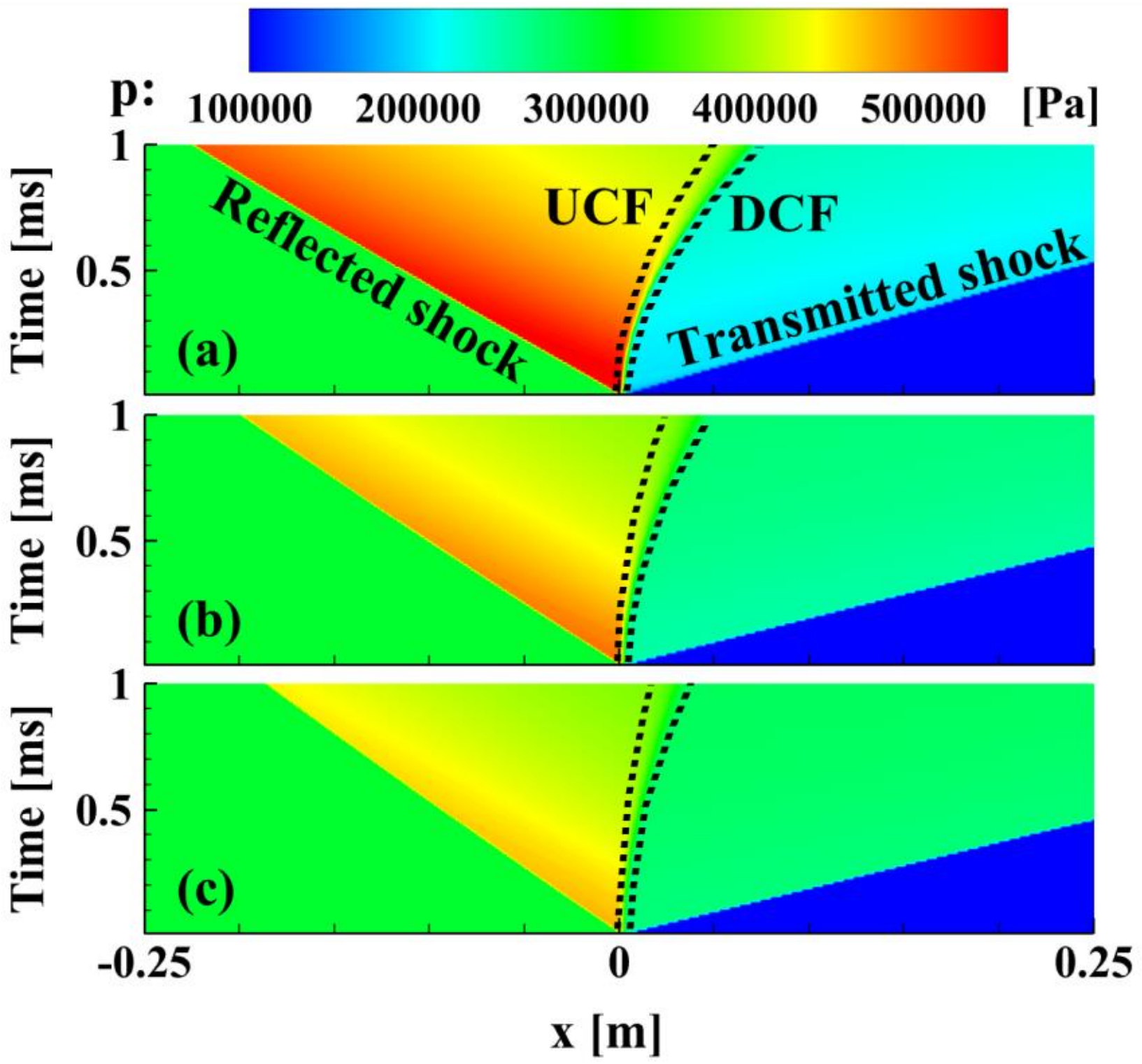

Fig. 13 $x$-$t$ diagram of gas pressure with different particle sizes: (a) 50, (b) 200, and (c) 400 μm.

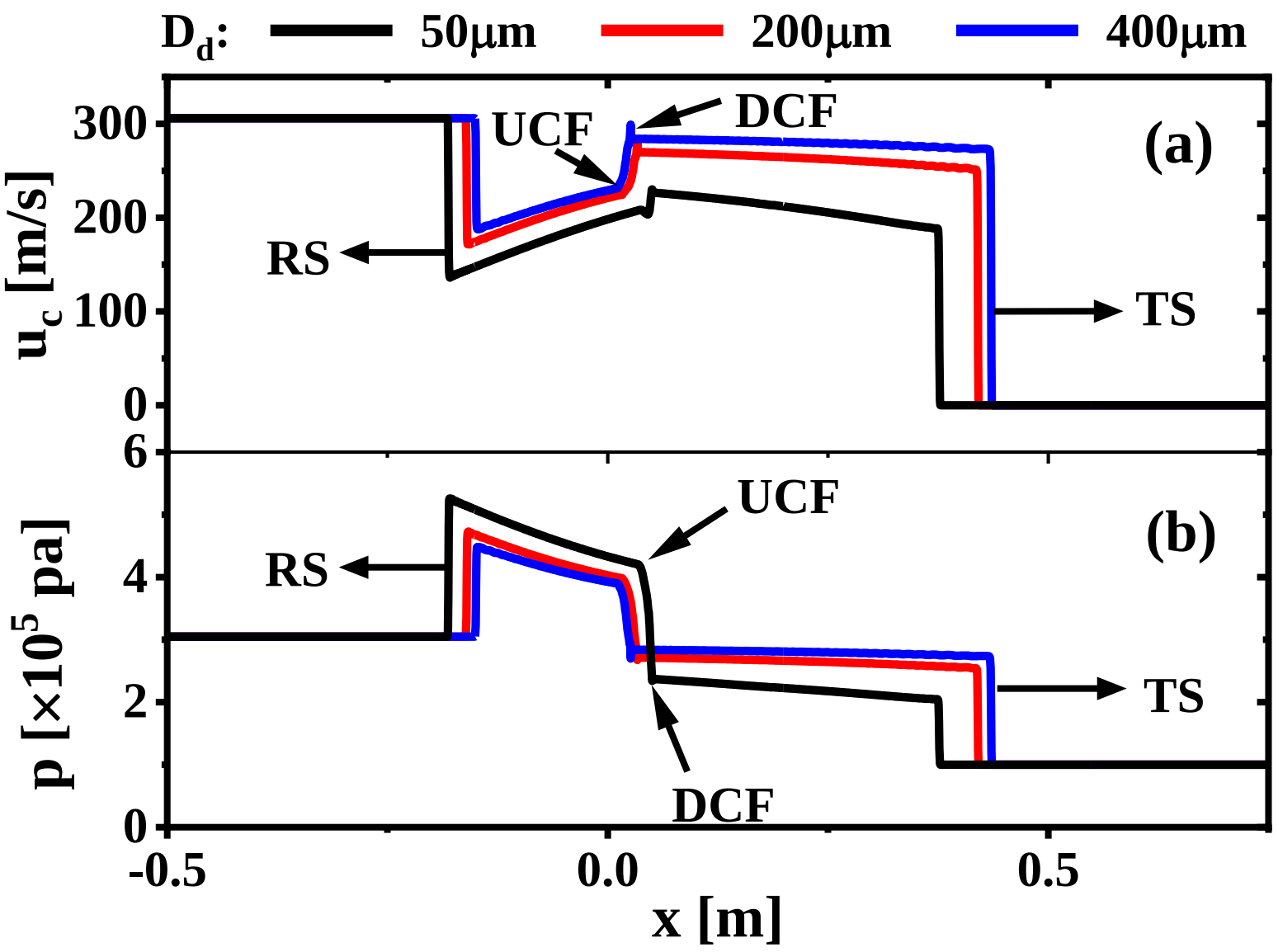

Fig. 14 Distributions of (a) gas velocity and (b) pressure with different particle diameters. RS: reflected shock; TS: transmitted shock. Data from 800 μs in Fig. 13.

Figure 13 shows the corresponding pressure evolutions in the $x$-$t$ diagram, whilst Fig. 14 presents the spatial profiles of the gas velocity and pressure at 800 μs. When the particle volume fraction is $\alpha_d$ = 10%, the acoustic impedance mismatch across the curtain interfaces, i.e., UCF and DCF, is irrespective of the particle size. Nonetheless, due to the inertia of the particles at the front, the two-phase mixture cannot respond instantly to the shocked gas motion. Thus, the equivalent gas model is not valid and can only provide an estimation. This is the reason why the interfacial gas dynamics are different when the droplet size varies in Fig. 13. Specifically, for $D_d$ = 50 μm, the reflected (transmitted) shock is the strongest (weakest). When the incident shock interacts with particle curtain, smaller particles (with larger specific area) lead to larger drag force on the carrier gas. This means that the incident shock is subject to larger resistance from the curtain. Thus, weaker shock transmits through the curtain, with a stronger reflected shock. Compared with the fluid dynamic force, the collision force exerted on the leading front particles, i.e., UCF and DCF, are much smaller. Meanwhile, inter-particle collision force does not directly interact with the carrier gas, and thus has almost negligible effects on shock waves.

As the particle diameter increases from 50 to 400 μm, the reflected (transmitted) shock intensity decreases (increases), see Fig. 13. Meanwhile, the reflected (transmitted) shock speed decreases (increases), as shown in Fig. 14. As for the gas velocity (see Fig. 14a), decreasing particle size leads to smaller average velocity across the curtain, i.e., between the UCF and DCF, because small particles accelerate faster and extract more kinetic energies from the gas phase. This is also consistent with the distributions of gas kinetic energy revealed by Jiang and Li [20].

## 6.2 Bi-dispersed particle curtain

### 6.2.1 Particle size effects

In section 6.1, it is found that the particle collision effects are generally of secondary importance when the mono-sized particles are considered. In this section, the particle collision effects will be further studied in modelling bi-dispersed particle curtains. Four cases are selected, and they are listed in Table 2. M1 is the base case, with mono-dispersed particles of $D_d$ = 200 μm, which has been studied in section 6.1. In M2−M4, two particle diameters are considered, and the larger and smaller particles have equal volume fraction (i.e., 5%) and material parameters, e.g., $\rho_d$ and $c_{p,d}$. In the following analysis, we use "group" to term the equally sized particles in the curtain. Two groups of particles are uniformly mixed in the curtain.

Table 2 Information of bi-dispersed particle curtain cases.

| $D_d$ [μm] | M1 | M2 | M3 | M4 |
|---|---|---|---|---|
| 25 | - | 5% | - | - |
| 50 | - | - | 5% | - |
| 100 | - | - | - | 5% |
| 200 | 10% | 5% | 5% | 5% |

The UCF and DCF trajectories of different particle groups in cases M2-M4 are shown in Fig. 15. In M2, the 25 μm UCF (DCF) particle moves faster than the 200 μm UCF (DCF) one. Two groups are ultimately separated at 1.8 ms (annotated with the dashed line in Fig. 15a). After separation, the 25 μm UCF move ahead of the 200 μm DCF, and hence the direct interactions between two groups do not exist. Meanwhile, the distance between the two groups continuously increases.

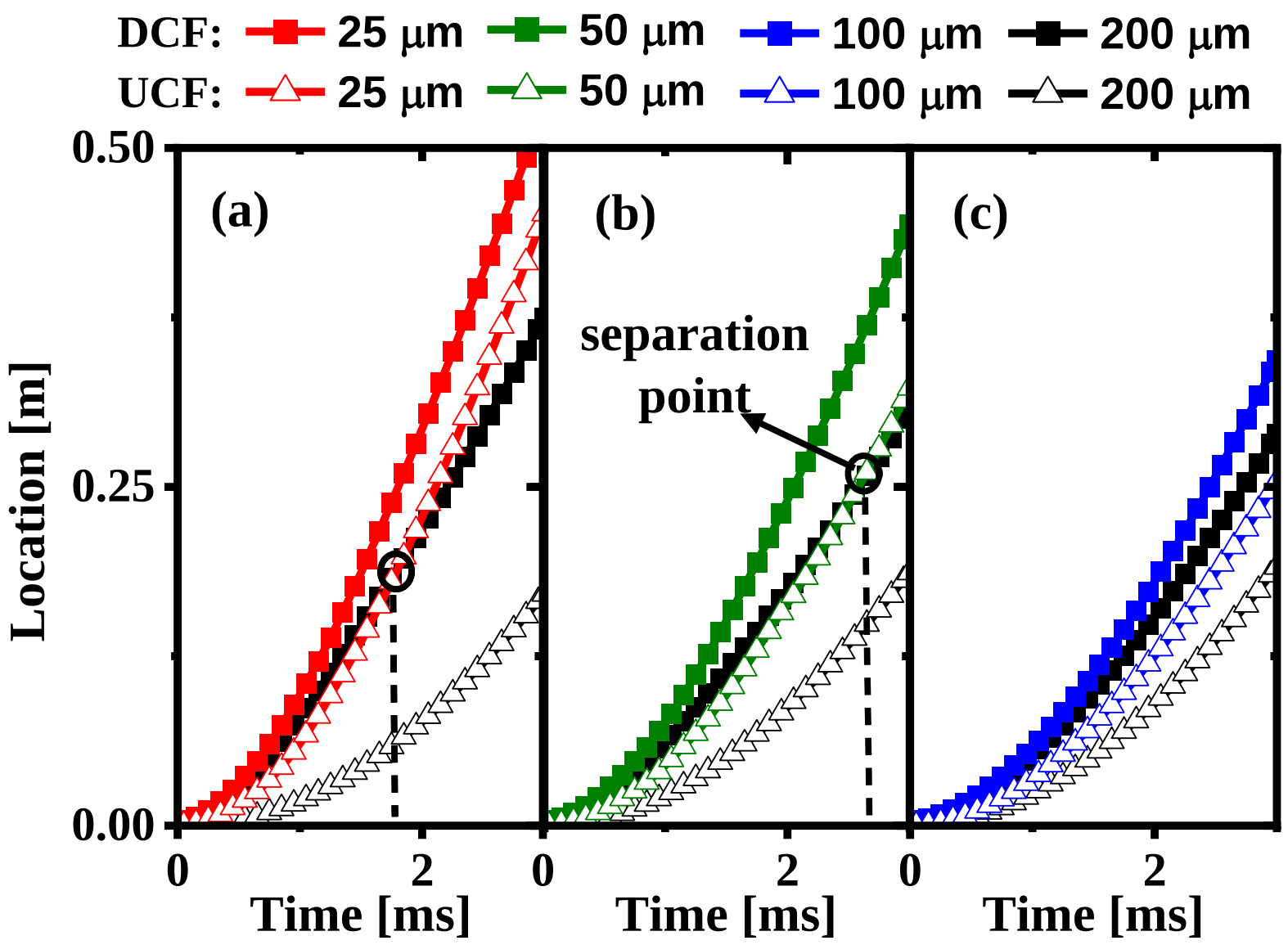

Fig. 15 Time evolutions of the UCF and DCF: (a) M2, (b) M3, and (c) M4.

Since the 200 μm DCF and 25 μm UCF particles have direct interactions before the separation point, here we only show the evolutions of the force exerted on them in Fig. 16(a) and 17(a), respectively. As shown in Fig. 16(a), the collision force on the 200 μm DCF particle in M2 mainly exists within the first 0.6 ms, beyond which it is around zero. This is because when $t$ > 0.6 ms, the particle volume fraction $\alpha_d$ drops to below 2%, as can be found in Fig. 16(b). Subsequently, the flow becomes dilute and thus the particle collision is almost negligible. Moreover, one can see from Fig. 17(a) that, for the 25 μm UCF particle, the $\mathbf{F}_{\mathrm{pp}}$ is negative at the very beginning but close to zero later, because the local particle volume fraction $\alpha_d$ drops to below 2%, see Fig. 17(b). Here the negative $\mathbf{F}_{\mathrm{pp}}$ means the 25 μm UCF particle has the resistance to its motion from the collision. This corresponds to the case 2 in

Table *1*.

In the bi-dispersed particle curtain, when the smaller particle diameter increases from 25 to 50 μm, the separation is delayed to 2.7 ms in M3, as can be found in Fig. 15(b). This is caused by slower motion of the 50 μm UCF. From Fig. 17(a), although $\mathbf{F}_{\mathrm{surf}}$ of the 50 μm UCF particle is larger than that of the 25 μm one, the resulting acceleration of the former is still smaller than the latter due to the mass difference, which has been explained in section 6.1.2.

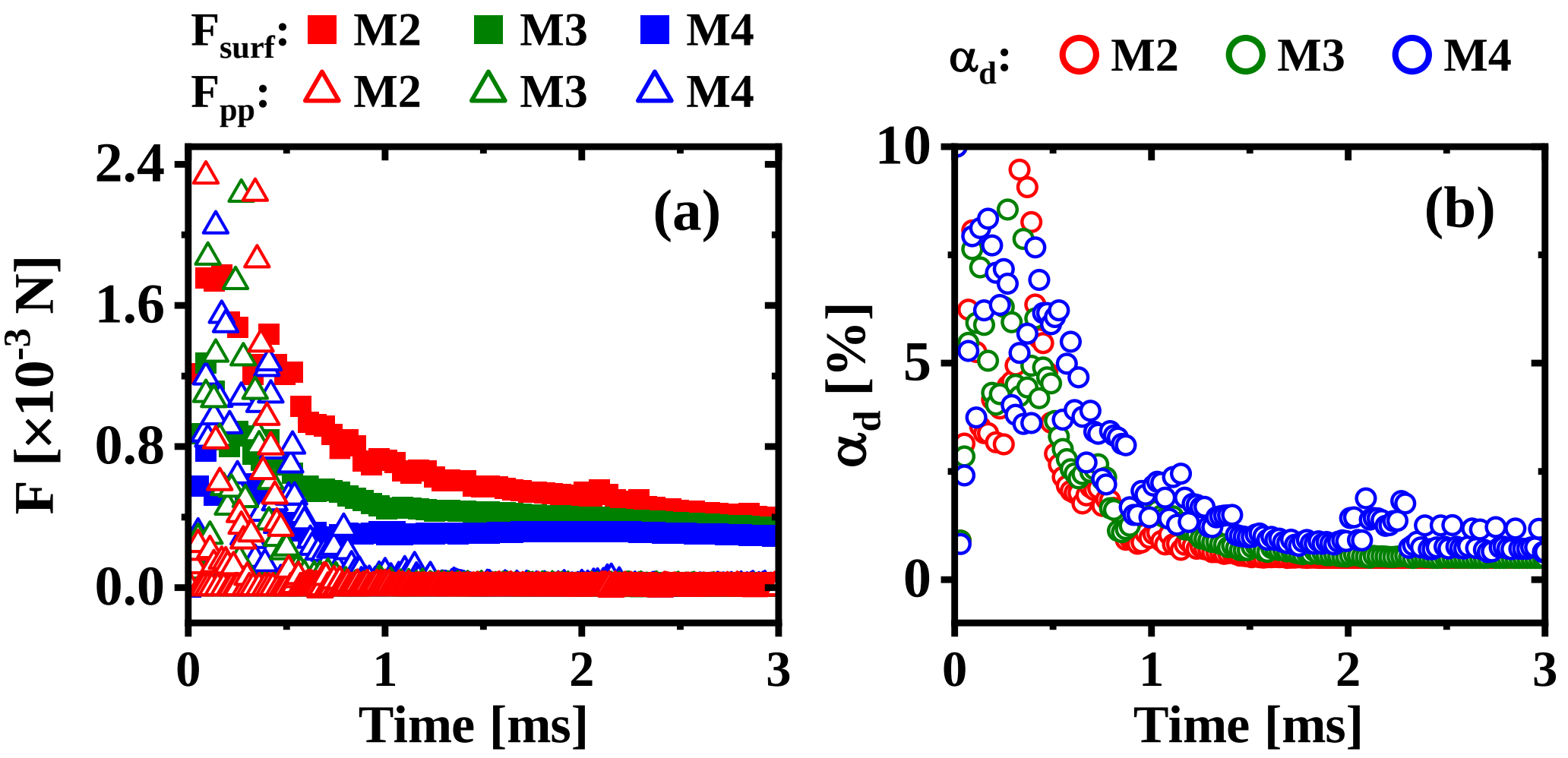

Fig. 16 Time evolutions of (a) the force on the 200 μm DCF particles and (b) particle volume fraction of the local CFD cell.

In M4, separation of two groups is not observed during the simulation time. However, such mixing does not cause obvious collision effects on the 200 μm DCF particle when $t$ > 1.2 ms, because of the low $\alpha_d$ (below 2%, see Fig. 16b). As for the displacement of 200 μm UCF particles in three cases, it marginally increases when the smaller particles increase from 25 to 100 μm. The 200 μm UCF particles almost do not have the collision force, similar to the results in Fig. 7 and Fig. 11.

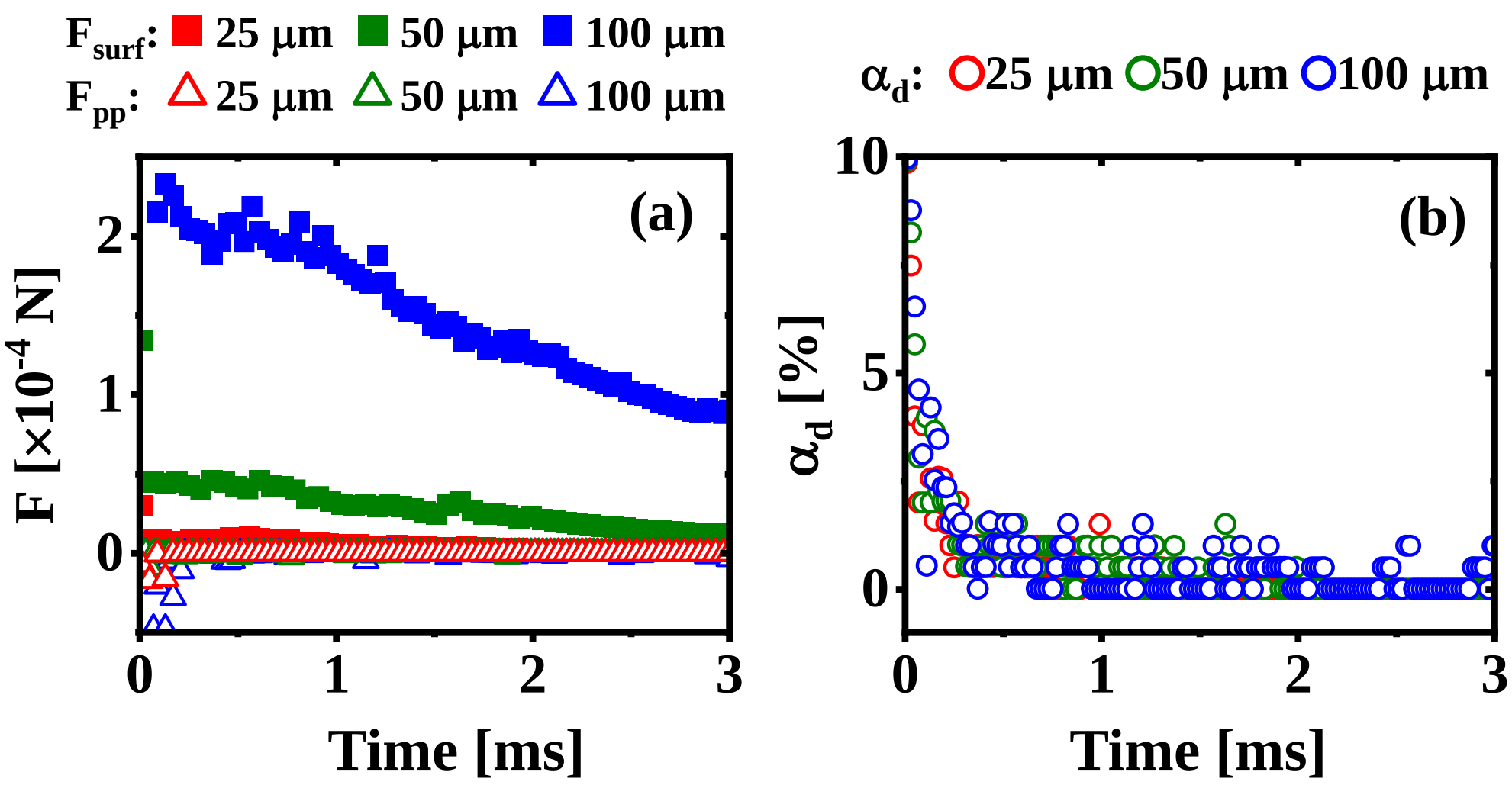

Fig. 17 Time evolutions of (a) the UCF particle force of small particle group, (b) particle volume fraction of the local CFD cell.

To elaborate on the particle collision effects on different particle groups, we further analyse the particle velocities and forces of case M3 at 10 μs in Fig. 18. The size ($\Delta x$) of each CFD cell is 500 μm, and the cell centre corresponds to a dashed line. The mass-averaged particle velocity $\tilde{\mathbf{u}}_\mathrm{d}$ in each cell is denoted with a blue circle at the cell centre.

Take the cell of 1.0 − 1.5 mm in Fig. 18(a) as an example. The $\tilde{\mathbf{u}}_\mathrm{d}$ is between the large particle and small particle velocities. This is reasonable because the large particles generally have smaller velocities, and vice versa. Figures 18(c) and 18(d) show the fluid dynamic force and collision force of two particle groups, respectively. In Fig. 18(c), the upstream particles generally have larger $\mathbf{F}_\mathrm{surf}$ than the downstream ones in each group, which is caused by the larger velocity difference ($\mathbf{u}_\mathrm{c} - \mathbf{u}_\mathrm{d}$) in the upstream zone, see Fig. 18(a). As shown in Fig. 18(d), the 50 μm particles at the left side of the cell centre are subject to negative $\mathbf{F}_\mathrm{pp}$, whilst the 200 μm particles at the right side of cell centre have positive $\mathbf{F}_{pp}$. The former particles correspond to case 2 whilst the latter ones correspond to case 3, as detailed in section 2.3. Conversely, the 50 μm particles at the right side and 200 μm particles at the left side have no collision force, which correspond to case 1 and case 4 in section 2.3, respectively. It can be found that particle collisions mainly happen between different particle groups with large velocity difference. For comparison, Fig. 18(b) presents the particle and gas velocity of M3 with the collision model turned off. One can see that no direct interactions between two groups can be observed. Then the velocity distribution of a particle group is more relatively continuous.

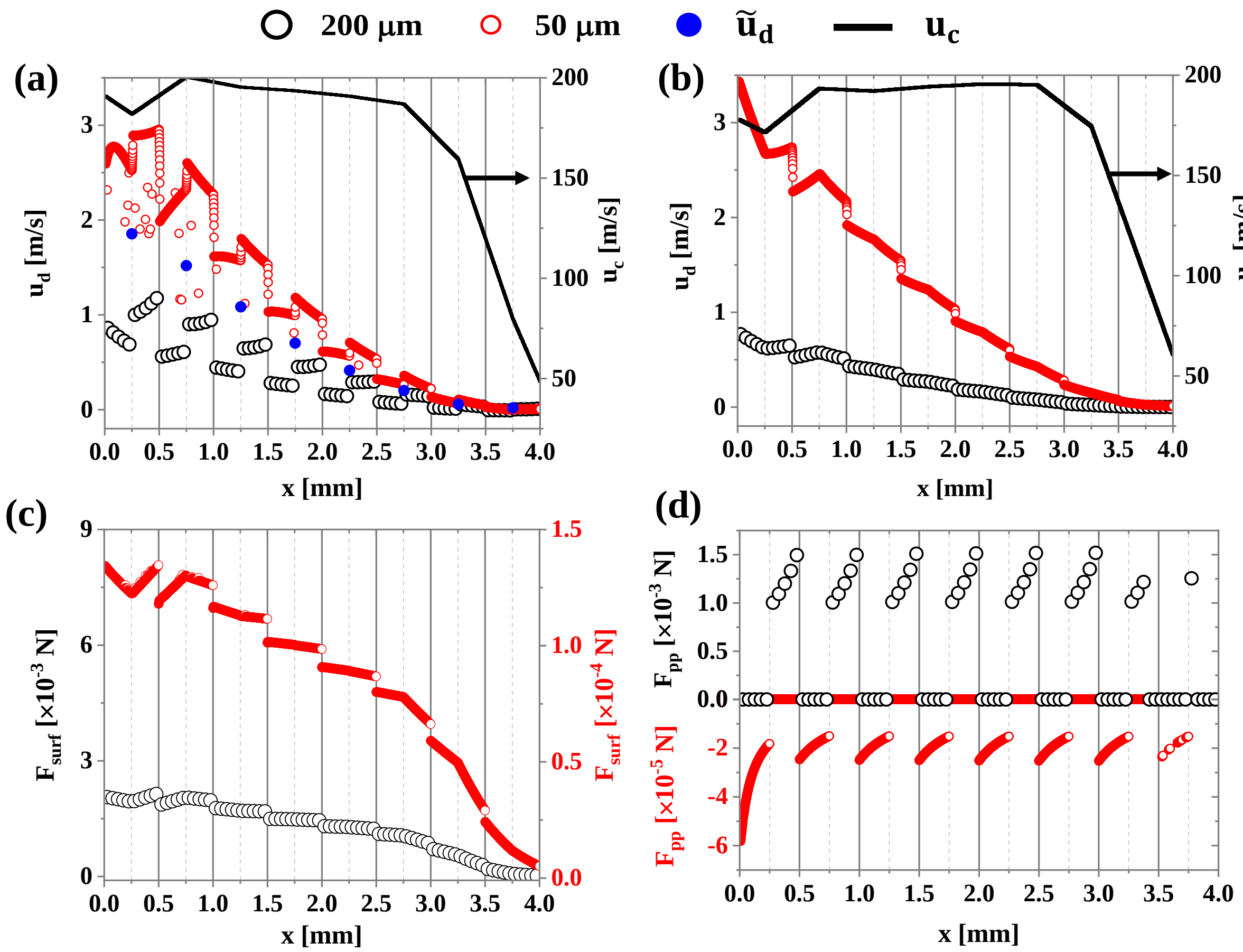

Fig. 18 Particle velocity distributions in M3 at 10 μs with the MPPIC collision model (a) on and (b) off. Particle force distributions of (c) fluid dynamic force and (d) collision force.

The distributions of the particle velocity considering collision effects are further discussed with Fig. 19(a). At 6 μs, the upstream particles accelerate first due to earlier contact with the incident shock. The 50 μm (200 μm) particles respond faster (slower) to the background flows and thus have larger (smaller) velocities. At 10 μs, scattering of the particle velocity can be observed in both groups, and at 100 μs such phenomenon becomes more obvious. This is caused by the particle collision effects between the two particle groups. Through the particle collision between different groups, some smaller particles decelerate while some larger particles accelerate, and thus resulting large velocity difference inside one group.

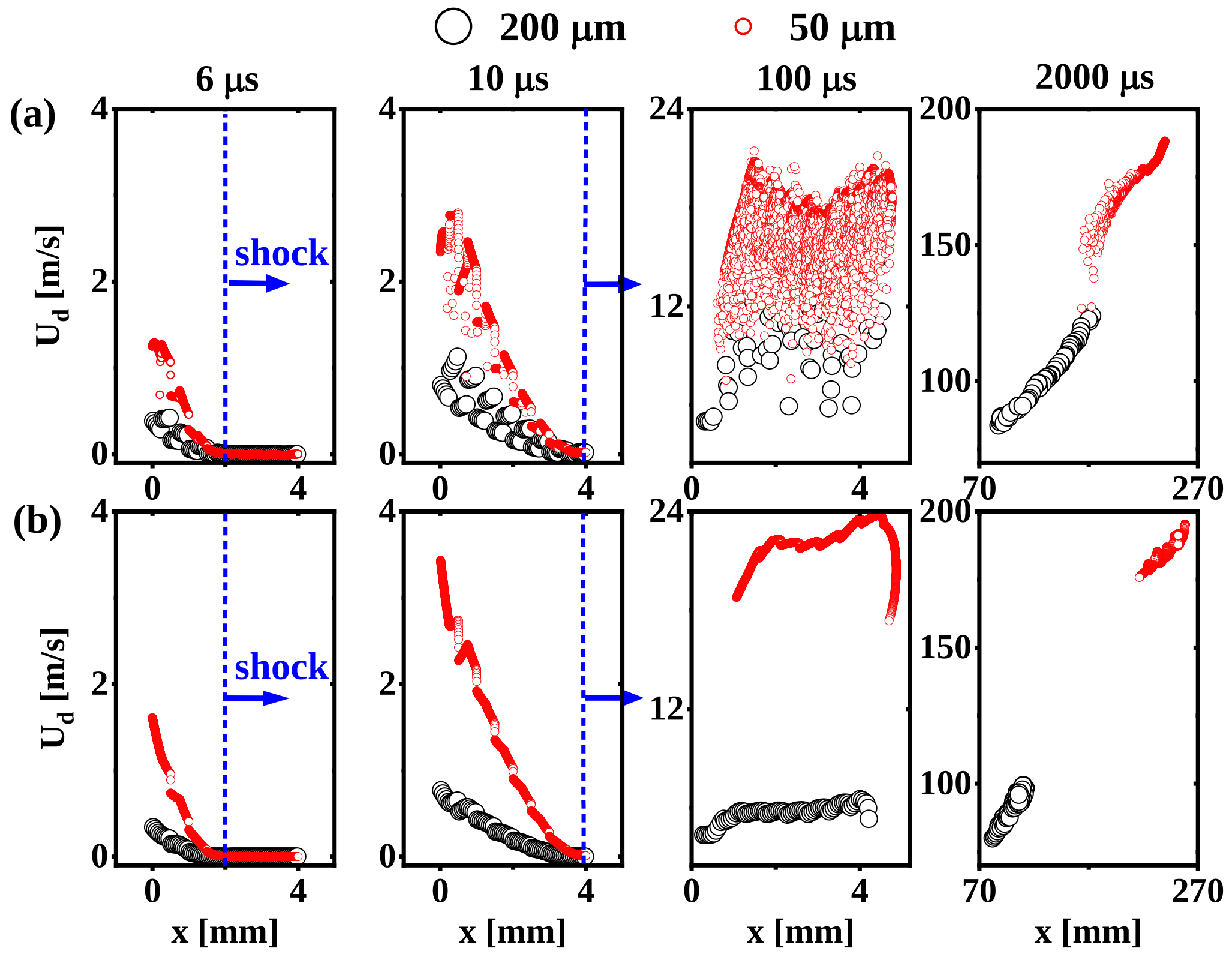

Fig. 19 Spatial distributions of particle velocities in case M3 at four instants: (a) with the collision model, (b) without the collision model. Dashed line: shock wave.

At 2,000 μs, group separation can be observed, and the smaller particles run faster (hence ahead) than the larger ones. Meanwhile, in both groups, the downstream particles have larger velocity than the upstream ones. This is because after the shock leaves the curtain, positive gas velocity gradient exists inside the curtain. Higher velocity difference exists between the gas and downstream particles, leading to larger $\mathbf{F}_{surf}$ on them. As a result, the downstream particles move faster than the upstream ones. However, this would not result in collisions of neighboring particles, as the particles correspond to *cases 1* and *4* in section 2.3.

Figure 19(b) shows the counterpart results when the collision model is turned off. It can be found that two groups of particles move without direct interactions and velocity fluctuations are limited. Compared with Fig. 19(a), the velocities of the 25 μm (200 μm) particles are generally over-predicted (under-predicted) when particle collision is not considered, because the momentum transfer from 25 to 200 μm particles due to collision effects is absent. At 2,000 μs, two particle groups are completely separated. When the collision effects are excluded, both groups have smaller thickness than that in Fig. 19(a). This is reasonable, as the collisions typically cause some small particles to decelerate while some large particles accelerate, as revealed from Figs. 15 and 18.

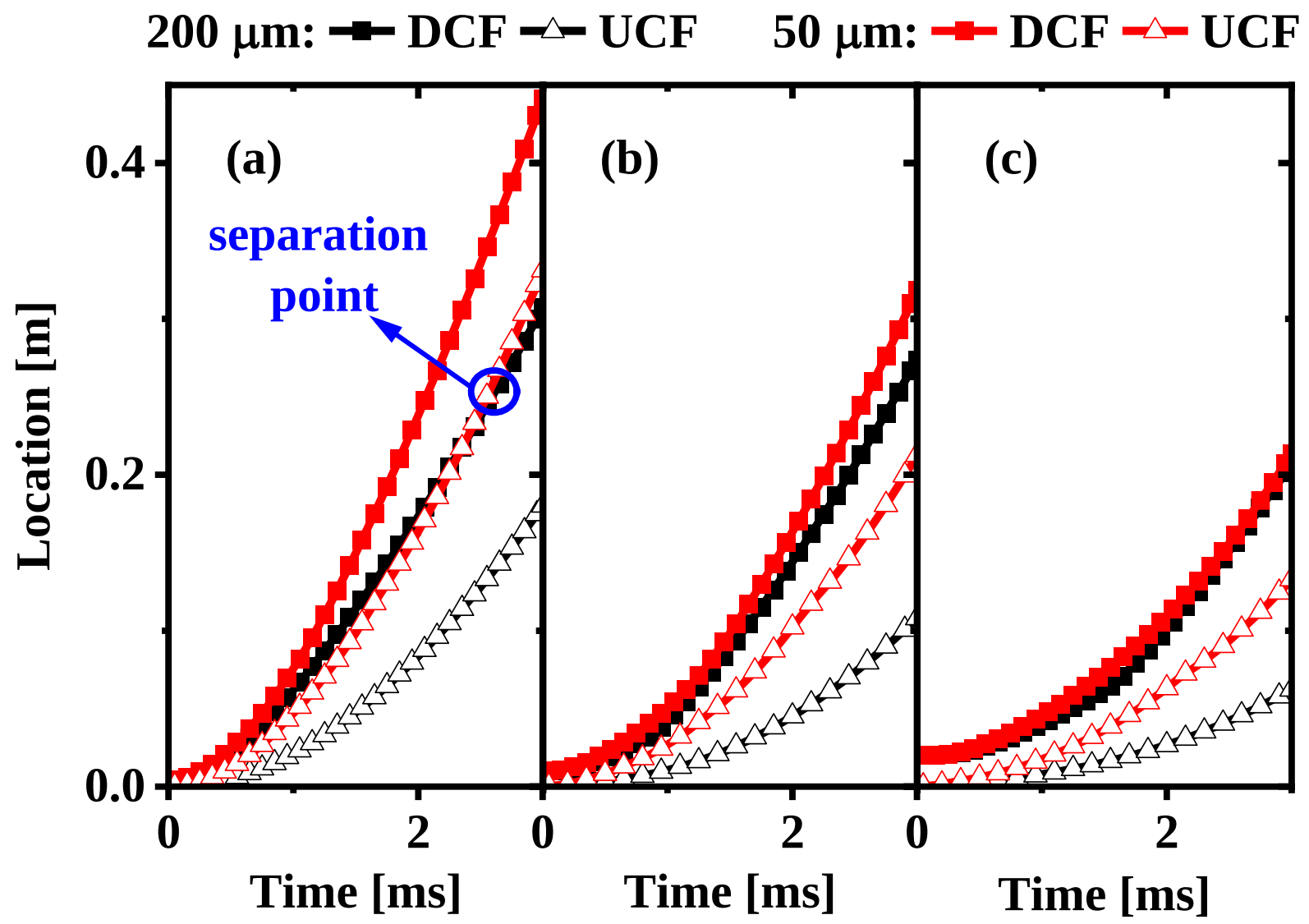

Fig. 20 Time evolutions of upstream and downstream curtain fronts with different curtain thickness: (a) L4, (b) L10, and (c) L20.

### 6.2.2 Curtain thickness effects

In the preceding discussion, the curtain thickness is fixed to be 4 mm. In this section, three curtain thickness will be studied, i.e., 4, 10, and 20 mm. They are termed as L4, L10 and L20, respectively. Particles of $D_d$ = 50 and 200 μm are uniformly mixed, and each group account for 5% by volume. Figure 20 shows the UCF and DCF trajectories of two groups. The separation of 200 μm DCF and 50 μm UCF is only observed in L4. Particularly, two groups are inter-penetrating nearly all the time in L20. Generally, when the curtain thickness increases, all the DCFs and UCFs move slower. This is mainly because thicker curtain has larger inertia and thus responds slower to the surrounding gas.

Meanwhile, the inter-particle collisions also have non-negligible effects on the motion of leading front particles, e.g., 200 μm DCF, which will be detailed in Fig. 21. Numerical experiment has also been undertaken for L10 with the collision model turned off, as shown in Section E of the supplementary documentary. One can see that the collision effects would dramatically delay the separation of two particle groups.

For simplicity, Fig. 21 only shows the particle forces and volume fractions (in the local cell) of the 200 μm DCF in three cases. In L20, there are two stages within which collisions can be observed, i.e., $0 - 0.5$ ms and $1.7 - 2.8$ ms. This peculiar phenomenon can be explained from Fig. 21(b): the particle volume fractions of these two periods are high. Specifically, the first collision stage is mainly because the two particle groups are not separated, e.g., 0.25 ms in Fig. *21*(c). The local particle volume fraction $\alpha_d$ of the 200 μm DCF particle is large and hence collisions happen. For the second collision stage, we show the volume fraction distribution of one typical instant, 2 ms, in Fig. 21(c). The 200 μm DCF particle runs into the high-volume region composed of 50 μm particles, which leads to strong collision effects. Particularly, the collision force $\mathbf{F}_{\mathrm{pp}}$ is much higher than the fluid dynamic force $\mathbf{F}_{\mathrm{surf}}$ during this stage, which indicates that the collisions effects have larger impacts on the 200 μm DCF particle than the two-phase interactions. Besides, two peaks of the fluid dynamic force $\mathbf{F}_{\mathrm{surf}}$ are also observed in the two collision periods. This is because the drag force $\mathbf{F}_{\mathrm{qs}}$ is positively correlated with $\alpha_{\mathrm{d}}$, which can be found from Eq. (18).

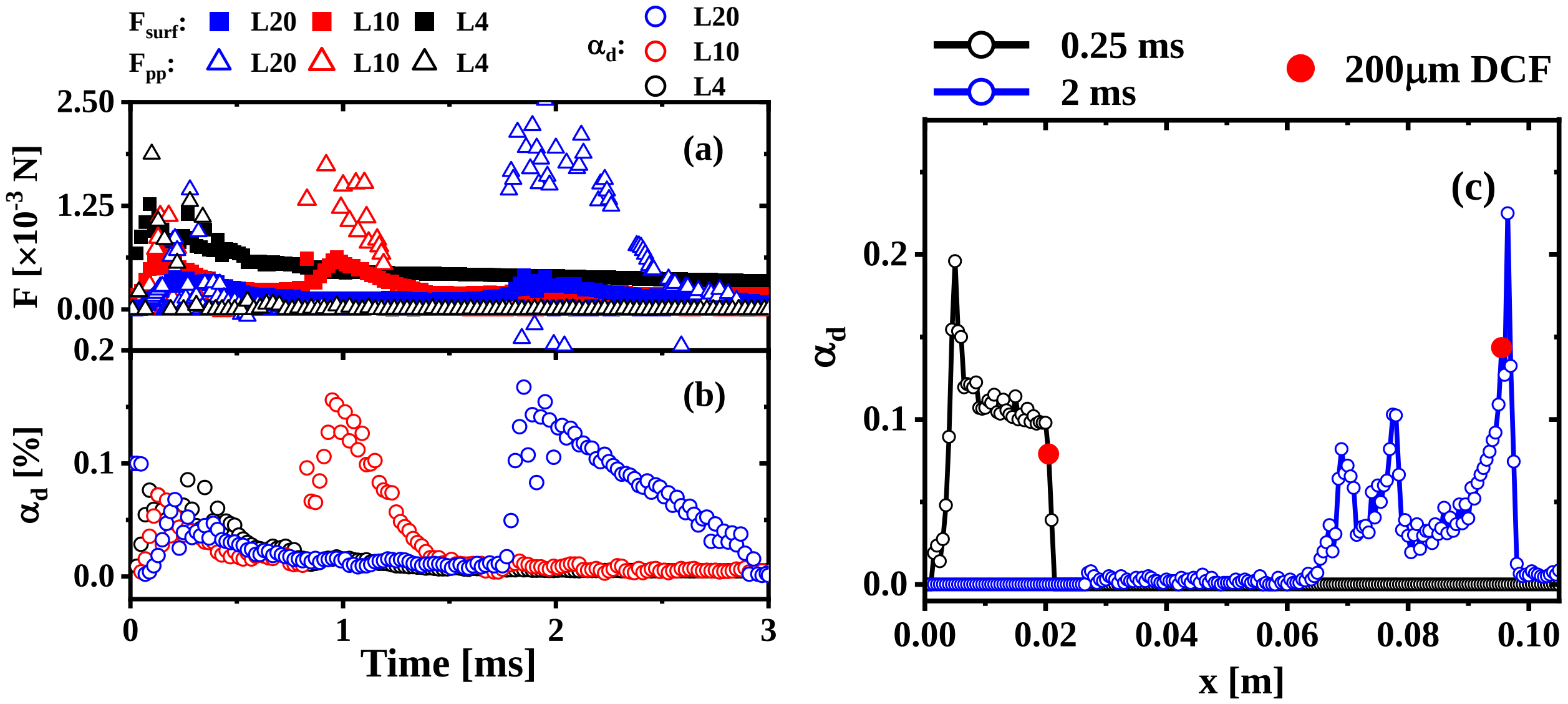

Fig. 21 Time evolutions of the 200 μm DCF particles: (a) force and (b) volume fraction in the local CFD cell; (c) spatial distribution of particle volume fraction in L20.

When the curtain thickness decreases to 10 mm, the second period of particle accumulation advances to 0.8 – 1.2 ms, and the peaks of $\mathbf{F}_{\mathrm{surf}}$ and $\mathbf{F}_{\mathrm{pp}}$ also appear earlier. When the curtain thickness further decreases to 4 mm, only the first accumulation period is observable, during which the collisions happen. This has been detailed in section 6.2.1.

Figure 22 shows the evolutions of the particle velocity distributions with different curtain thickness. The shock wave leaves the 20 mm curtain at 0.05 ms. In L20, two particle groups are mixed for almost all the time, as demonstrated in Fig. 22(a). From the analysis in section 6.2.1, such mixing would cause strong collision effects between them. Figure 23 shows the spatial profiles of the collision force of two groups and the particle volume fractions at 2 ms. Generally, the 200 μm particles have positive $\mathbf{F}_{\mathrm{pp}}$, while the 50 μm particles negative ones. Obviously, in L20, three peaks of volume fraction correspond to three peaks of $\mathbf{F}_{\mathrm{pp}}$ in 200 μm group, but only one peak of $\mathbf{F}_{\mathrm{pp}}$ in 50 μm group. The is because small particles move faster and most accumulate in the largest peak volume, i.e., the peak near 0.1 m. Moreover, relatively weaker collision effects can be observed in the regions between different peaks, i.e., the relatively dilute regime.

When the initial curtain thickness decreases, the group separation becomes earlier, see Fig. 22. Besides, the whole curtain thickness increases at a later time, e.g., 1 and 2 ms in Figs 22(b) and 22(c).

As a result, the spatial distributions of the particle volume fraction and collision forces decrease in Fig. 23. Specifically, only one weaker peak of $\alpha_d$ is observed in L10, but none in L4 at 2 ms. This is because smaller initial curtain thickness with larger expansion rate would generally result in lower $\alpha_d$, and thus no local particle accumulation to induce possible particle collisions.

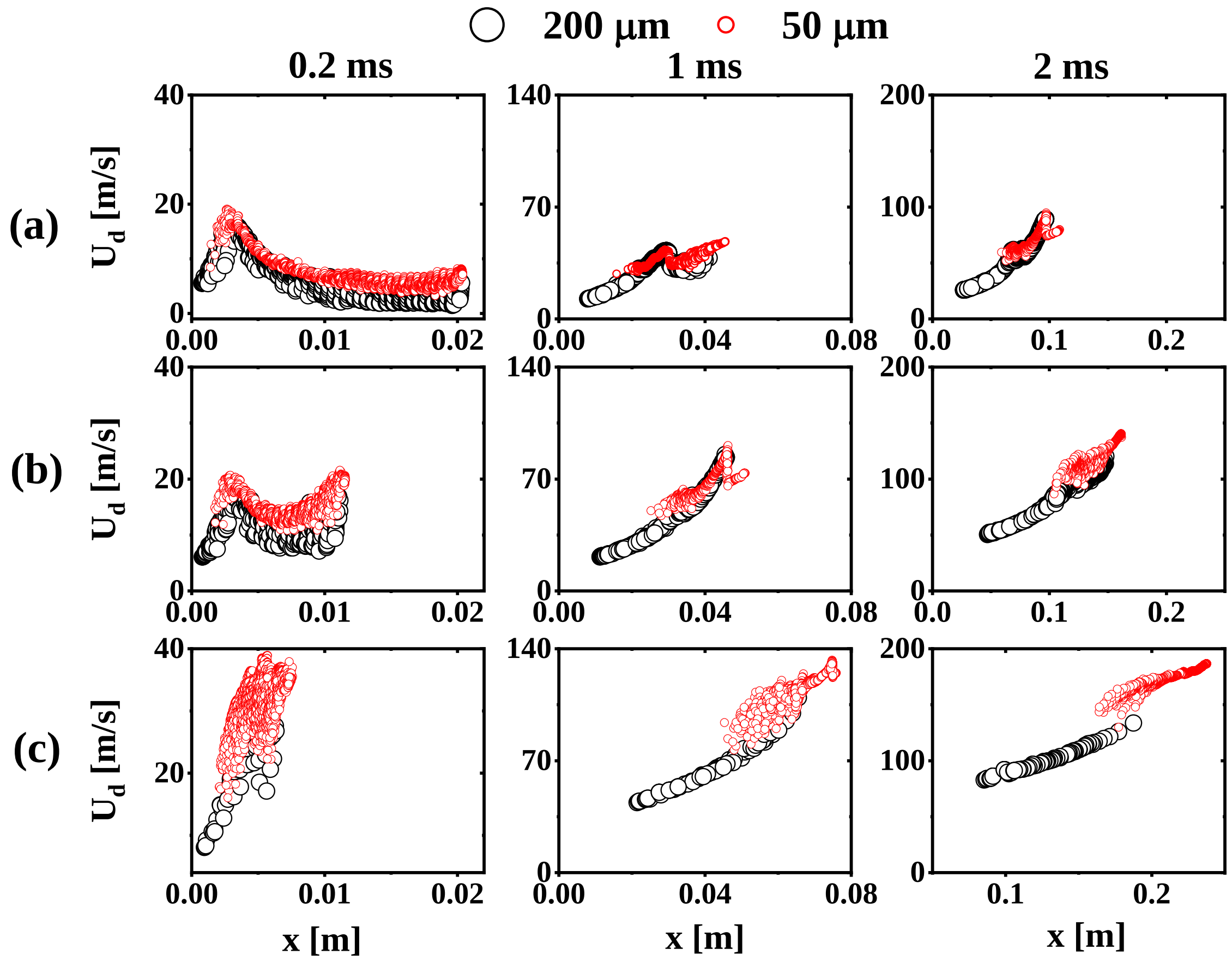

Fig. 22 Spatial distributions of particle velocities at three instants: (a) L20, (b) L10, and (c) L4.

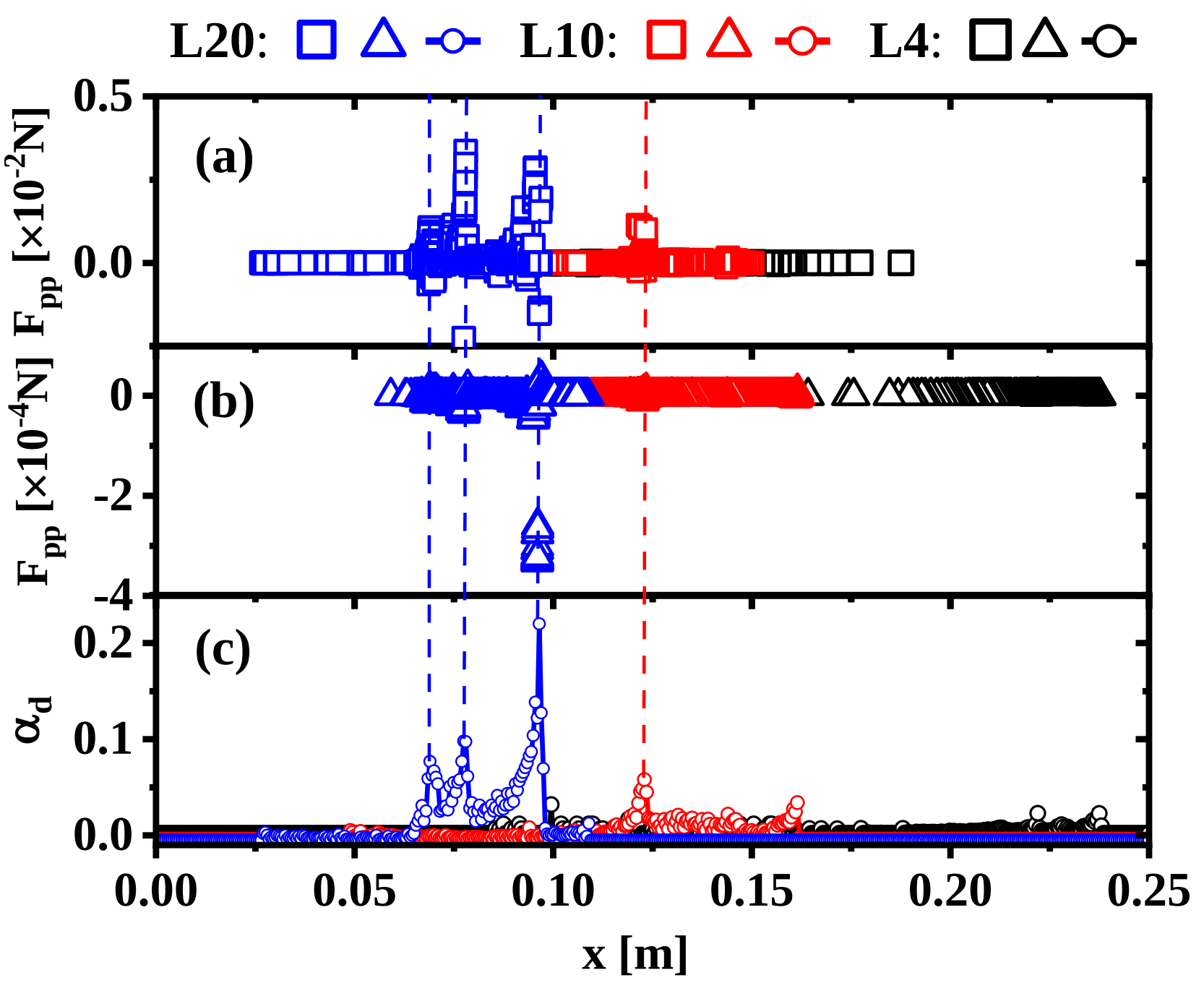

Fig. 23 Spatial distributions of particle collision forces in (a) 200 µm and (b) 50 µm groups, and (c) particle volume fractions at 2 ms.

## 7. CONCLUSIONS

In this study, the interactions between an incident shock and a moderately dense particle curtain are numerically investigated. Based on the extended *RYrhoCentralFoam* solver with the MP-PIC model and an improved drag model, compressible gas-particle flows ranging from dilute to moderately dense regime can be modelled under the Eulerian-Lagrangian framework. Two validation cases suggest that the solver is accurate to reproduce the movement of particle curtain and the evolution of gas phase quantities. Based on our analysis, the following main conclusions can be drawn:

(1) Particle volume fraction has larger effect on UCF than DCF due to the differentiated fluid dynamic force. After the constant-thickness regime, it is the varying-acceleration regime rather than the constant-acceleration regime. A larger particle volume fraction leads to a stronger reflected shock and a weaker transmitted shock due to a higher difference of acoustic impedence across the curtain front.

(2) Small particle curtains have faster moving curtain fronts. When the particle size decreases to 50 µm, a short compression period can be observed before curtain expansion. Smaller particles lead to a stronger reflected shock and a weaker transmitted shock due to the larger drag force to the gas phase.

(3) In the bi-dispersed particle curtain, the collision effect is significant in the region where two-particle groups co-exist. The collision decelerates small particles while accelerates large particles, causing the scattering of particle velocity. When the collision model is turned off, the momentum transfer from small to large particles is absent.

(4) When the thickness of bi-dispersed particle curtain is 4 mm, only one collision stage can be observed for 200 μm DCF. However, when the thickness increases to 10 and 20 mm, two collision stages appear due to the delayed separation of two particle groups. Furthermore, the delayed separation results in strong collision effects in the multiple high-volume peaks of mixing regions. When two groups are separated, the collision effects are diminished.

The above findings indicate that under moderately dense conditions, the collision model may be unnecessary for predicting the curtain morphological evolutions in mono-dispersed particles. However, in bi-dispersed or poly-dispersed particles, the collision effects are important and must be modelled.

## ACKNOWLEFDGEMENT

The computational work for this article was performed on the resources from the National Supercomputing Center, Singapore (https://www.nscc.sg/). PZ is supported by the NUSRI-CQ Research Scholarship.

## DATA AVAILABILITY

The data that support the findings of this study are available from the corresponding author upon reasonable request.